%% file: feinbergzee.tex
\documentstyle[12pt]{article}
\input epsfig

\begin{document}
\baselineskip=20.5pt
\def\beqra{\begin{eqnarray}} \def\eeqra{\end{eqnarray}}
\def\beqast{\begin{eqnarray*}} \def\eeqast{\end{eqnarray*}}
\def\beq{\begin{equation}}      \def\eeq{\end{equation}}
\def\be{\begin{enumerate}}   \def\ee{\end{enumerate}}

\def\fnote#1#2{\begingroup\def\thefootnote{#1}\footnote{#2}\addtocounter
{footnote}{-1}\endgroup}

\def\ut#1#2{\hfill{UTTG-{#1}-{#2}}}
\def\fl#1#2{\hfill{FERMILAB-PUB-94/{#1}-{#2}}}
\def\itp#1#2{\hfill{NSF-ITP-{#1}-{#2}}}

\def\gam{\gamma}
\def\Gam{\Gamma}
\def\la{\lambda}
\def\eps{\epsilon}
\def\La{\Lambda}
\def\si{\sigma}
\def\Si{\Sigma}
\def\al{\alpha}
\def\Tha{\Theta}
\def\tha{\theta}
\def\vphi{\varphi}
\def\del{\delta}
\def\Del{\Delta}
\def\ab{\alpha\beta}
\def\om{\omega}
\def\Om{\Omega}
\def\mn{\mu\nu}
\def\mun{^{\mu}{}_{\nu}}
\def\kap{\kappa}
\def\rsi{\rho\sigma}
\def\beal{\beta\alpha}
\def\til{\tilde}
\def\rta{\rightarrow}
\def\eqv{\equiv}
\def\nab{\nabla}
\def\pa{\partial}
\def\sit{\tilde\sigma}
\def\ul{\underline}
\def\indt{\parindent2.5em}
\def\nd{\noindent}
\def\rsi{\rho\sigma}
\def\beal{\beta\alpha}
\def\caa{{\cal A}}
\def\cb{{\cal B}}
\def\cac{{\cal C}}
\def\cd{{\cal D}}
\def\ce{{\cal E}}
\def\cf{{\cal F}}
\def\cg{{\cal G}}
\def\cah{{\cal H}}
\def\ci{{\cal I}}
\def\cj{{\cal{J}}}
\def\ck{{\cal K}}
\def\cl{{\cal L}}
\def\cm{{\cal M}}
\def\cn{{\cal N}}
\def\cO{{\cal O}}
\def\cp{{\cal P}}
\def\car{{\cal R}}
\def\cs{{\cal S}}
\def\ct{{\cal{T}}}
\def\cu{{\cal{U}}}
\def\cv{{\cal{V}}}
\def\cw{{\cal{W}}}
\def\cx{{\cal{X}}}
\def\cy{{\cal{Y}}}
\def\cz{{\cal{Z}}}
\def\asymptotic{{_{\stackrel{\displaystyle\longrightarrow}
{x\rightarrow\pm\infty}}\,\, }} 
\def\asymptext{\raisebox{.6ex}{${_{\stackrel{\displaystyle\longrightarrow}
{x\rightarrow\pm\infty}}\,\, }$}} 

\def\raisenot{\raise .5mm\hbox{/}}
\def\nota{\ \hbox{{$a$}\kern-.49em\hbox{/}}}
\def\notA{\hbox{{$A$}\kern-.54em\hbox{\raisenot}}}
\def\notb{\ \hbox{{$b$}\kern-.47em\hbox{/}}}
\def\notB{\ \hbox{{$B$}\kern-.60em\hbox{\raisenot}}}
\def\notc{\ \hbox{{$c$}\kern-.45em\hbox{/}}}
\def\notd{\ \hbox{{$d$}\kern-.53em\hbox{/}}}
\def\notbd{\ \hbox{{$D$}\kern-.61em\hbox{\raisenot}}} 
\def\note{\ \hbox{{$e$}\kern-.47em\hbox{/}}}
\def\notk{\ \hbox{{$k$}\kern-.51em\hbox{/}}}
\def\notp{\ \hbox{{$p$}\kern-.43em\hbox{/}}}
\def\notq{\ \hbox{{$q$}\kern-.47em\hbox{/}}}
\def\notW{\ \hbox{{$W$}\kern-.75em\hbox{\raisenot}}}
\def\notz{\ \hbox{{$Z$}\kern-.61em\hbox{\raisenot}}}
\def\notpa{\hbox{{$\partial$}\kern-.54em\hbox{\raisenot}}}
\def\fo{\hbox{{1}\kern-.25em\hbox{l}}}  
\def\rf#1{$^{#1}$}
\def\bx{\Box}
\def\tr{{\rm Tr}}
\def\rmtr{{\rm tr}}
\def\dgg{\dagger}
\def\lag{\langle}
\def\rag{\rangle}
\def\bmid{\big|}
\def\vlap{\overrightarrow{\La p}} 
\def\lrta{\longrightarrow} \def\lrar{\raisebox{.8ex}{$\longrightarrow$}}
\def\ON{{\cal O}(N)}
\def\UN{{\cal U}(N)}
\def\bdPh{\mbox{\boldmath{$\dot{\!\Phi}$}}}
\def\bPh{\mbox{\boldmath{$\Phi$}}}
\def\bPhs{\bPh^2}
\def\sef{S_{eff}[\sigma,\pi]}
\def\sigx{\sigma(x)}
\def\pix{\pi(x)}
\def\bph{\mbox{\boldmath{$\phi$}}}
\def\bphs{\bph^2}
\def\ex{\BM{x}}
\def\exs{\ex^2}
\def\xdot{\dot{\!\ex}}
\def\y{\BM{y}}
\def\ys{\y^2}
\def\ydot{\dot{\!\y}}
\def\pat{\pa_t}
\def\pax{\pa_x}

\renewcommand{\thesection}{\arabic{section}}
\renewcommand{\theequation}{\thesection.\arabic{equation}}

\begin{flushright}
\itp{96}{15}\\
March 26, 1996\\
\hfill{hep-th/9603173}\\
\end{flushright}

{\em To the memory of}~ {\bf R.\,F. Dashen} $1938-1995$

\vspace*{.3in}
\begin{center}
  \Large{\sc Dynamical Generation of Extended Objects in a $1+1$ 
Dimensional Chiral Field Theory: Non-Perturbative Dirac Operator 
Resolvent Analysis} 
\normalsize

\vspace{36pt}
{\large Joshua Feinberg\fnote{*}{{\it e-mail: joshua@itp.ucsb.edu}}
 \& A. Zee}\\
\vspace{12pt}
 {\small \em Institute for Theoretical Physics,}\\ {\small \em
University of California, Santa Barbara, CA 93106, USA}
\vspace{.6cm}

\end{center}

\begin{minipage}{5.3in}
{\abstract~~~~~
We analyze the $1+1$ dimensional Nambu-Jona-Lasinio model 
non-perturbatively. In addition to its simple ground state saddle 
points, the effective action of this model has a rich collection of 
non-trivial saddle points in which the composite fields 
$\sigx=\lag\bar\psi\psi\rag$ and $\pix=\lag\bar\psi i\gam_5\psi\rag$ 
form static space dependent configurations because of non-trivial 
dynamics. These configurations may be 
viewed as one dimensional chiral bags that trap the original fermions 
(``quarks") into stable extended entities (``hadrons"). We provide 
explicit expressions for the profiles of these objects and calculate 
their masses. Our analysis of these saddle points is based on an explicit 
representation we find for the diagonal resolvent of the Dirac 
operator in a $\{\sigx, \pix\}$ background which produces a prescribed 
number 
of bound states. We analyse in detail the cases of a single as well as 
two 
bound states. We find that bags that trap $N$ fermions are the most 
stable ones, because they release all the fermion rest mass as binding 
energy and become massless. Our explicit construction of the diagonal 
resolvent 
is based on elementary Sturm-Liouville theory and simple dimensional 
analysis and does not depend on the large $N$ approximation. 
These facts make it, in our view, simpler and more direct than the 
calculations previously done by Shei, using the inverse scattering method 
following Dashen, Hasslacher, and Neveu. Our method of finding such 
non-trivial static configurations may be applied to other $1+1$ dimensional 
field theories.}

\end{minipage}

\vspace{48pt}

\vfill
\pagebreak

\setcounter{page}{1}

\section{Introduction}

Over the last thirty years or so, physicists have gradually learned about
the behavior of quantum field theory in the non-perturbative regime. In
$1+1$ dimensional spacetime, some models are exactly soluble 
\cite{rajaraman}. Another important approach involves the large $N$ 
expansion \cite{brezin}. In particular, in the
mid-seventies Dashen, Hasslacher, and Neveu \cite{dhn} used the inverse
scattering method \cite{faddeev} to determine the spectrum of the 
Gross-Neveu model \cite{gn}. 
Recently, one of us developed an alternative method, based on the
Gel'fand-Dikii equation \cite{gd}, to study the same problem \cite{josh1} 
as well as other problems \cite{josh2}.
As will be explained below, we feel that this method has certain 
advantages over the inverse scattering method.

In this paper we study the $1+1$ dimensional Nambu-Jona-Lasinio 
(NJL)~\footnote{This model is also dubbed in the literature as the 
``Chiral Gross-Neveu Model" as well as the ``Multiflavor Thirring 
Model".} model\cite{njl} which is a renormalisable field theory defined 
by 
the action\cite{gn}
\beq
S=\int d^2x\left\{\sum_{a=1}^N\, \bar\psi_a\,i\notpa\,\psi_a +
\frac{g^2}{2}\; \left[\left( \sum_{a=1}^N\;\bar\psi_a\,\psi_a\right)^2
-\left(\sum_{a=1}^N\;\bar\psi_a\gam_5\psi_a\right)^2\right]\right\}
\label{lagrangian}
\eeq
describing $N$ self interacting massless Dirac fermions 
$\psi_a\,(a=1,\ldots,N)$. This action is invariant under $SU(N)_f\otimes 
U(1)\otimes U(1)_A$, namely, under
\beqra
\psi_a &\rightarrow & U_{ab} \psi_b\,,\quad\quad U\in 
SU(N)_f\,,\nonumber\\
\psi_a &\rightarrow & e^{i\alpha}\psi_a\,,\nonumber\\
{\rm and}\quad\quad \psi_a &\rightarrow & e^{i\gam_5\beta}\psi_a\,.
\label{symmetries}
\eeqra

We rewrite (\ref{lagrangian}) as
\beq
S=\int 
d^2x\,\left\{\bar\psi\left[i\notpa-(\si+i\pi\gam_5)\right]\psi-{1\over 
2g^2}\,(\si^2+\pi^2)\right\}
\label{auxiliary}
\eeq
where $\si(x), \pi(x)$ are the scalar and pseudoscalar auxiliary fields, 
respectively \footnote{From this point to the end of this paper flavor 
indices
are suppressed.  Thus $i\bar\psi\notpa\psi$ should be understood as
$\displaystyle{i\sum^N_{a=1} \bar\psi_a \notpa\psi_a}$.  Similarly
$\bar\psi\Gamma\psi$ stands for $\displaystyle{\sum^N_{a=1} 
\,\bar\psi_a\Gamma\psi_a}$, where $\Gamma=1,\gam_5$.}, which are both of 
mass dimension $1$. These 
fields
are singlets under $SU(N)_f\otimes U(1)$, but transform as a vector under 
the axial transformation in (\ref{symmetries}), namely
\beq
\si+i\gam_5\pi\rightarrow e^{-2i\gam_5\beta}(\si + i\gam_5\pi)\,.
\label{vector}
\eeq
 
Thus, the partition function associated with (\ref{auxiliary}) is
\beq
\cz=\int\,\cd\si\,\cd\pi\,\cd\bar\psi\,\cd\psi \,\exp i\, 
\int\,d^2x\left\{\bar\psi
\left[i\notpa-\left(\si+i\pi\gam_5\right)\right]\psi-{1\over 
2g^2}\,\left(\si^2+\pi^2\right)\right\}
\label{partition}
\eeq
Integrating over the grassmannian variables leads to 
$\cz=\int\,\cd\si\,\cd\pi\,\exp \{iS_{eff}[\si,\pi]\}$
where the bare effective action is
\beq
S_{eff}[\si,\pi] =-{1\over 2g^2}\int\, d^2x 
\,\left(\si^2+\pi^2\right)-iN\, 
\tr\ln\left[i\notpa-\left(\si+i\pi\gam_5\right)\right]
\label{effective}
\eeq
and the trace is taken over both functional and Dirac indices. 

This theory has been studied in the limit 
$N\rightarrow\infty$ with $Ng^2$ held fixed\cite{gn}. In this limit 
(\ref{partition}) is governed by saddle points of (\ref{effective}) 
and the small fluctuations around them. The most general saddle point
condition reads

\beqra
{\del S_{\em eff}\over \del \si\left(x,t\right)}  &=&
-{\si\left(x,t\right)\over g^2} + iN ~{\rm tr} \left[~~~~~ \langle x,t | 
{1\over i\notpa
-\left(\si + i\pi\gam_5\right)} | x,t \rangle \right]= 0
\nonumber\\{}\nonumber\\
{\del S_{\em eff}\over \del \pi\left(x,t\right)}  &=&
-{\pi\left(x,t\right)\over g^2} - ~N~ {\rm tr} \left[~\gam_5~\langle x,t 
| {1\over i\notpa
-\left(\si + i\pi\gam_5\right)} | x,t \rangle~\right] = 0\,.
\label{saddle}
\eeqra

In particular, the non-perturbative vacuum of (\ref{lagrangian}) is 
governed 
by the simplest large $N$ saddle points of the path integral associated 
with 
it, where the composite scalar operator $\bar\psi\psi$ and the 
pseudoscalar operator $i\bar\psi\gam_5\psi$ develop space time 
independent expectation values.  

These saddle points are extrema of the effective potential $V_{eff}$ 
associated with (\ref{auxiliary}), namely, the value of  $-S_{eff}$ for 
space-time independent $\si,\pi$ configurations per unit time per unit 
length.  The effective potential $V_{eff}$ 
depends only on the combination $\rho^2=\si^2+\pi^2$ as a result of 
chiral symmetry. $V_{eff}$ has a minimum as a function of $\rho$ at  
$\rho = m \neq 0$ 
that is fixed by the (bare) gap equation\cite{gn}
\beq
-m + iNg^2\,{\rm tr}\int
{d^2k\over\left(2\pi\right)^2}{1\over\notk-m}
= 0
\label{bgap}
\eeq
which yields the dynamical mass
\beq
m = \Lambda\,e^{-{\pi\over Ng^2\left(\Lambda\right)}}\,.
\label{mass}
\eeq
Here $ \Lambda$ is an ultraviolet cutoff. The mass $m$ must be a 
renormalisation group invariant. Thus, the model is asymptotically 
free. We can get rid of the cutoff at the price of introducing an 
arbitrary 
renormalisation scale $\mu$. The renormalised
coupling $g_R\left(\mu\right)$ and the cut-off dependent bare
coupling are then related through $ \Lambda\,e^{-{\pi\over 
Ng^2\left(\Lambda\right)}} = 
\mu\,e^{1-{\pi\over Ng_R^2\left(\mu\right)}}$ in a convention where 
$Ng_R^2\left(m\right) = {1\over\pi}$. Trading the dimensionless coupling 
$g_R^2$ for the dynamical mass scale $m$ represents the well known 
phenomenon of dimensional transmutation. 

The vacuum manifold of (\ref{auxiliary}) is therefore a circle 
$\rho=m$ in the $\si,\pi$ plane, and the equivalent vacua are 
parametrised by the chiral angle $\theta={\rm arctan} {\pi\over\sigma}$. 
Therefore, small fluctuations of the Dirac fields around the vacuum 
manifold 
develop dynamical\footnote{Note that the axial 
$U(1)$ symmetry in (\ref{symmetries}) protects the fermions from 
developing a mass term to any order in perturbation theory.}. 
chiral mass $m\,{\rm exp} (i\theta \gam_5)$.

Note in passing that the massless fluctuations of $\theta$ along the 
vacuum manifold decouple from the spectrum \cite{decouple}
so that the axial $U(1)$ symmetry does not break dynamically in this two 
dimensional model \cite{coleman}.

Non-trivial excitations of the vacuum, on the other hand, are described 
semiclassically by large $N$ saddle points of the path integral over 
(\ref{lagrangian}) at which $\si$ and $\pi$ develop space-time dependent 
expectation values\cite{cjt,jg}. These expectation values are the 
space-time dependent solution of (\ref{saddle}). Saddle points of this 
type are important 
also in discussing the large order behavior\cite{bh,dev} of the 
${1\over N}$ expansion of the path integral over (\ref{lagrangian}).

Shei \cite{shei} has studied the saddle points of the NJL model by 
applying
the inverse scattering method following Dashen et al.\cite{dhn}. 
These saddle points describe sectors of (\ref{lagrangian}) that include 
scattering states of the (dynamically massive) fermions in 
(\ref{lagrangian}), as well as a rich collection of bound states thereof. 

These bound states result from the strong infrared interactions, 
which polarise the vacuum inhomogeneously, causing 
the composite scalar $\bar\psi\psi$ and pseudoscalar 
$i\bar\psi\gam_5\psi$ 
fileds to form finite action space-time dependent condensates. These 
condensates are stable becuse of the binding energy released by the 
trapped fermions and therefore cannot form without such binding. This
description agrees with the general physical picture drawn in 
\cite{mackenzie}. 
We may regard these condensates as one dimensional chiral 
bags \cite{sphericalbag,shellbag} that trap the original fermions 
(``quarks") into 
stable finite action extended entities (``hadrons").

In this paper we develop further the method of \cite{josh1,josh2}, 
applying it to 
the NJL model (\ref{lagrangian}) as an alternative to the inverse 
scattering investigations in \cite{shei}. We focus on static extended 
configurations 
providing explicit expressions for the profiles of these objects and 
calculate their masses.  Our analysis of these static saddle points is 
based on an explicit representation we find for the diagonal resolvent of 
the Dirac operator in a $\sigx, \pix$ background which produces a 
prescribed number of bound states. This explicit construction of the 
diagonal resolvent can actually be carried out for finite $N$. 
It is based on elementary Sturm-Liouville theory as well as on simple 
dimensional analysis. All our manipulations involve the space dependent 
scalar and pseudoscalar condensates
directly. In our view, these facts make the method presented here simpler 
than inverse scattering calculations previously employed in this problem
because we do not need to work with the scattering data and the so 
called trace identities that relate them to the space dependent 
condensates.
Our method of finding such non-trivial static configurations may be 
applied to other two dimensional field theories.

It is worth mentioning at this point that the NJL model 
(\ref{lagrangian})
is completely integrable for any number of flavors\footnote{For $N=1$ a 
simple Fierz transformation shows that (\ref{lagrangian}) is simply the 
massless Thirring model, which is a conformal quantum field theory having 
no mass gap.
A mass gap appears dynamically only for $N\geq 2$.}~$N$. Its spectrum and 
completely factorised $S$ matrix were determined in a series of papers 
\cite{andrei} by a Bethe ansatz diagonalisation of the Hamiltonian 
for any number $N$ of flavors. The large $N$ spectrum obtained here as 
well as in \cite{shei} is consistent with the exact solution of 
\cite{andrei}.  Note, however, that the large $N$ analysis in this paper 
concerns only dynamics of 
the interactions between fermions and extended objects. We do not address 
issues like scattering of one extended object on another, which is 
discussed in the exact analysis of \cite{andrei}. Consistency of our 
approximate large $N$ results and the exact results of \cite{andrei} 
reassures us of the validity of our calculations. 

Rather than treating non-trivial excitations as abstract vectors in 
Hilbert space, which is inevitable in \cite{andrei}, our analysis draws 
almost a ``mechanical" picture of how 
``hadrons" arise in the 
NJL model. This description of ``hadron" formation as a result of 
inhomogeneous polarisations of the vacuum due to strong infrared 
interactions may have some restricted similarity to dynamics of QCD in 
the 
real world. Furthermore, our resolvent method is potentially applicable 
for non-integrable models 
in $1+1$ dimensions. In contrast, Bethe ansatz and factorisable $S$ 
matrix 
techniques are limited in principle to $1+1$ dimensions because of the 
Coleman-Mandula theorem\cite{colemanmandula}, whereas large $N$ 
saddle 
point techniques may provide powerful tools in analysis of more realistic 
higher 
dimensional field theories\cite{dhn1}.

If we set $\pix$ in (\ref{auxiliary}) to be identically zero, we recover 
the Gross-Neveu model, defined by
\beq
S_{GN}=\int d^2x\,\left\{\bar\psi\left[i\notpa-\si\right]\psi-{\si^2 
\over 
2g^2}\right\}\,.
\label{GN}
\eeq

In spite of their similarities, these two field theories are quite
different, as is well-known from the field theoretic literature of the
seventies. The crucial difference is that the Gross-Neveu model 
possesses a discrete symmetry, $\sigma\rightarrow -\sigma$, rather than 
the 
continuous symmetry (\ref{symmetries}) in the NJL model studied here. 
This discrete 
symmetry
is dynamically broken by the non-perturbative vacuum, and thus there is a 
kink solution \cite{ccgz,dhn,josh1}, the so-called 
Callan-Coleman-Gross-Zee (CCGZ) kink $\sigx = 
m\,{\rm tanh}(mx)$, interpolating between $\pm m$ at  $x= \pm \infty$ 
respectively. Therefore, topology insures the stability of these kinks. 

In contrast, the NJL model, with its continuous symmetry, does 
not have a topologically stable soliton solution. The  
solitons arising
in the NJL model and studied in this paper can only be stabilised by 
binding fermions. 
To stress this 
observation further, we note that the spectrum of the Dirac operator of 
the 
Gross-Neveu model in the backgound of a CCGZ kink has a single bound 
state 
at zero energy, and therefore no binding energy is released when they 
trap
(any number of) fermions. The stability of the kinks in the Gross-Neveu 
model is guaranteed by topology already. In contrast, the stability of 
the extended objects studied here is not due to topology, but to 
dynamics.

This paper is organised as follows: In Section 2 we prove that the static
condenstates $\sigx$ and $\pix$ in (\ref{auxiliary}) must be such that 
the
resulting Dirac operator is reflectionless.  Our proof of this strong 
restriction on the Dirac operator involves basic field theoretic 
arguments 
and has nothing to do with the large $N$ approximation. 
We next show in Section 3 that if we fix in advance the number of bound 
states in the spectrum of the reflectionless Dirac operator, then simple 
dimensional analysis determines the diagonal resolvent of this operator 
explicitly in 
terms of the background fields and their derivatives. 
We then construct 
the resolvent assuming the backround 
fields support a single bound state in Subsection 4.1. We are able to 
determine the profile of the background 
fields up to a finite number of parameters: the relative chiral rotation 
of 
the two vacua at the two ends of the one dimensional space and the bound 
state energies. In Subection 4.2 we provide partial analysis of the case 
of 
two bound states.  We stress again that our construction of these 
background fields has nothing to do with the large $N$ approximation.

In order to determine these parameters we have to impose the saddle point
condition. We do so in Sections 5.1 amd 5.2. The relative chiral rotation 
of  the asymptotic vacua is proportional to the number of fermions 
trapped in the 
bound states, in accordance with \cite{goldstone}.

Some technical details are left to 
two appendices. In Appendix A we derive
the spatial asymptotic behavior of the static Dirac operator Green's 
function. In order to make our paper self contained we derive the 
Gel'fand-Dikii 
equation in Appendix B.

\pagebreak

\section{Absence of Reflections
in the Dirac Operator With Static Background Fields }
\setcounter{equation}{0}

As was explained in the introduction, we are interested in static space 
dependent solutions of the extremum condition on $S_{\it eff}$. To this 
end 
we need to invert the Dirac operator 

\beq
D\equiv\left[i\notpa-(\sigx+i\pix\gam_5)\right]
\label{dirac}
\eeq
in a given background of static field configurations $\sigx$ and $\pix$. 
In particular, we have to find the diagonal resolvent of (\ref{dirac}) in 
that 
background. The extremum condition on $S_{\it eff}$ relates this 
resolvent, which in principle is a complicated and generally unknown 
functional
of $\sigx$, $\pix$ and of their derivatives, to $\sigx$ and $\pix$ 
themselves. This complicated relation is the source of all difficuties 
that arise in any attempt to solve the model under consideration. It 
turns out, however, that basic field theoretic considerations, that are 
unrelated to the extremum condition, imply that (\ref{dirac}) must be 
reflectionless. This spectral
property of (\ref{dirac}) sets rather powerful restrictions on the static 
background fields $\sigx$ and $\pix$ which are allowed dynamically. In 
the 
next section we show how this special property of (\ref{dirac}) allows us 
to write explicit expressions for the resolvent in some restrictive 
cases, that 
are interesting enough from a physical point of view.

Inverting (\ref{dirac}) has nothing to do with the large $N$ 
approximation, 
and consequently our results in this section are valid for any value of 
$N$.

Here $\sigx$ and $\pix$ are our static background field configurations, 
for which we assume asymptotic behavior dictated by simple physical 
considerations. The overall energy deposited in any relevant 
static $\si, \pi$ configuration must be finite. Therefore these 
fields must approach constant vacuum asymptotic values, while their 
derivatives vanish asymptotically. Then the axial $U(1)$ symmetry implies 
that $\si^2 + \pi^2 \asymptext m^2$, where 
$m$ is the dynamically generated mass,  and therefore we arrive at the 
asymptotic boundary conditions for $\si$ and $\pi$, 
\beqra
&&\si\asymptotic m{\rm cos}\theta_{\pm}\quad\quad ,\quad\quad 
\si'\asymptotic 0 
\nonumber\\{}\nonumber\\
&&\pi\asymptotic m{\rm sin}\theta_{\pm}\quad\quad , 
\quad\quad \pi'\asymptotic 0
\label{boundaryconditions}
\eeqra
where $\theta_{\pm}$ are the asymptotic chiral alignment angles. Only the 
difference $\theta_+ - \theta_-$ is meaningful, of course, and henceforth 
we use the axial symmetry to set $\theta_- = 0$, such that $\si 
(-\infty)=m$ and $\pi (-\infty)=0$. We also omit the subscript from 
$\theta_+$
and denote it simply by $\theta$ from now on.
It is in the background of such fields that we wish to invert 
(\ref{dirac}).

In this paper we use the Majorana
representation  $\gam^0=\si_2\;,\; \gam^1=i\si_3$ and 
$\gam^5=-\gam^0\gam^1=\si_1$ for $\gam$ matrices. In this representation  
(\ref{dirac}) becomes

\beq
D =\left(\begin{array}{cc} -\pa_x - \si & -i\omega - i\pi \\{}&{}\\ 
i\omega- i\pi & \pa_x - \si\end{array}\right)\,.
\label{dirac1}
\eeq

Inverting (\ref{dirac1}) is achieved by solving  
\beq
\left(\begin{array}{cc} -\pa_x - \sigx & -i\omega - i\pix \\{}&{}\\ 
i\omega- i\pix & \pa_x - \sigx\end{array}\right)\cdot 
\left(\begin{array}{cc} a(x,y) &  b(x,y) \\{}&{}\\ c(x,y) & 
d(x,y)\end{array}\right)\,=\,-i{\bf 1}\del(x-y)
\label{greens}
\eeq
for the Green's function of (\ref{dirac1}) in a given background 
$\sigx,\pix$.
By dimensional analysis, we see that the quantities $a,b,c$ and 
$d$ are dimensionles.

The diagonal elements $a(x,y), ~d(x,y)$ in (\ref{greens}) may be 
expressed
in term of the off-diagonal elements as
\beq
a(x,y)={i\left[\pa_x-\sigx\right]c\left(x,y\right)\over 
\omega-\pix}\,,\quad\quad
d(x,y)={i\left[\pa_x+\sigx\right]b\left(x,y\right)\over \omega+\pix}
\label{ad}
\eeq
which in turn satisfy the second order partial differential equations
\beqra
&&-\pa_x\left[{\pa_x b(x,y)\over 
\omega+\pix}\right]+\left[\sigx^2+\pix^2-\si'(x)-\omega^2+{\sigx\pi'(x)
\over 
\omega+\pix}\right]{b(x,y)\over \omega+\pix}\,=\,~~\del(x-y)
\nonumber\\&&{}\nonumber\\
&&-\pa_x\left[{\pa_x c(x,y)\over 
\omega-\pix}\right]+\left[\sigx^2+\pix^2+\si'(x)-\omega^2+{\sigx\pi'(x)
\over 
\omega-\pix}\right]{c(x,y)\over 
\omega-\pix}\,=\,-\del(x-y)\,.\nonumber\\&&{}
\label{bc}
\eeqra

Thus, $b(x,y)$ and $-c(x,y)$ are simply the Green's functions of the 
corresponding second order Sturm-Liouville operators in (\ref{bc}),
\beqra
b(x,y)&=&{\theta\left(x-y\right)b_2(x)b_1(y)+\theta\left(y-x\right)b_2(y)
b_1(x) \over W_b}
\nonumber\\{}\nonumber\\
c(x,y)&=&-{\theta\left(x-y\right)c_2(x)c_1(y)+\theta\left(y-x\right)c_2(y
)c_1(x)\over W_c}\,.
\label{bcexpression}
\eeqra
Here $b_1(x)$ and $b_2(x)$ are the Jost functions of the first equation 
in (\ref{bc}) 
and  
\beq
W_b={b_2(x)b_1^{'}(x)-b_1(x)b_2^{'}(x)\over \omega+\pix}\,.
\label{wronskian}
\eeq
is their Wronskian. The latter is independent of $x$, since $b_1$ and 
$b_2$ share a common value  
of the  
spectral parameter $\omega^2$. Similarly, $c_1, c_2$ are the Jost 
functions 
of the second equattion in (\ref{bc}) and $W_c$ is their Wronskian.
We leave the precise definition of these Jost functions in terms of their 
spatial asymptotic behavior to Appendix A, where we also derive the 
spatial asymptotic behavior of the static Dirac operator Green's 
function.
Substituting (\ref{bcexpression}) into (\ref{ad}) we 
obtain
the appropriate expressions for $a(x,y)$ and $d(x,y)$, which we do not 
write explicitly.\footnote{It is useful however to note, that despite the 
$\pax$ operation in (\ref{ad}), neither $a(x,y)$ nor $d(x,y)$ contain 
pieces proportional to $\del(x-y)$\,. Such pieces cancel one another due 
to the symmetry of (\ref{bcexpression}) under $x\leftrightarrow y$\,.}

We define the diagonal resolvent 
$\langle x\,|iD^{-1} | x\,\rangle$ symmetrically as
\beqra
\langle x\,|-iD^{-1} | x\,\rangle &\equiv& \left(\begin{array}{cc} A(x) & 
B(x) \\{}&{}\\ C(x) & 
D(x)\end{array}\right)\nonumber\\{}\nonumber\\{}\nonumber\\
 &=& {1\over 2} \lim_{\epsilon\rightarrow 0+}\left(\begin{array}{cc} 
a(x,y) + a(y,x) &  b(x,y) + b(y,x)\\{}&{}\\ c(x,y) +c(y,x) & d(x,y) + 
d(y,x)\end{array}\right)_{y=x+\epsilon}\,.
\label{diagonal}
\eeqra
Here $A(x)$ through $D(x)$ stand for the entries of the diagonal 
resolvent, which following (\ref{ad}) and (\ref{bcexpression}) have the 
compact representation\footnote{$A, B, C$ and $D$ are obviously functions 
of $\omega$ as
well. For notational simplicity we suppress their explicit $\omega$ 
dependence.} 
\beqra
B(x)&=&~~{b_1(x)b_2(x)\over W_b}\quad\quad , \quad\quad D(x)={i\over 
2}{\left[\pax+2\sigx\right]B\left(x\right)\over 
\omega+\pix}\,,\nonumber\\
C(x)&=&-{c_1(x)c_2(x)\over W_c}\quad\quad , \quad\quad A(x)={i\over 
2}{\left[\pax-2\sigx\right]C\left(x\right)\over \omega-\pix}\,.
\label{abcd}
\eeqra

A simplifying observation is that the two linear operators on the left 
hand side of the equations (\ref{bc}) transform one into the other under 
a 
simultaneous sign flip\footnote{This is merely a reflection of the fact 
that coupling the fermions to $\pi\gam_5$ does not respect charge 
conjugation invariance.} of $\sigx$ and $\pix$.  Therefore  
$c(\si,\pi)=-b(-\si,-\pi)$\,, and in particular
\beq
C(\si,\pi)=-B(-\si,-\pi)\,, 
\label{bcrelation}
\eeq
and thus all four entries of the diagonal resolvent (\ref{diagonal}) may 
be expressed in terms of $B(x)$.

The spatial asymptotic behavior of (\ref{diagonal}) is derived in 
Appendix A and given by (\ref{asymptoticABCD}). A more compact form of 
that result
is 
\beqra
\langle x\,|-iD^{-1} | x\,\rangle \asymptotic &&{1 + 
R\left(k\right)e^{2ik\,|x|}\over 2k}
\left[i\gam_5\pix - \sigx - \gam^0\om\right] 
\nonumber\\{}\nonumber\\
&&+\,{R\left(k\right)e^{2ik\,|x|}\over 2}\,\gam^1~ {\rm sgn}\,x
\label{asymptoticdirac}
\eeqra
where $k=\sqrt{\om^2-m^2}$~ and $R(k)$ is the reflection coefficient
of the first equation in (\ref{bc}).

Note that for $\om^2>m^2$, {\em i.e.}, in the continuum part of the 
spectrum
of (\ref{dirac1}), the piece of the resolvent (\ref{asymptoticdirac}) 
that is
proportional to $R(k)$ oscillates persistently as a function of $x$. This 
observation has a far reaching result that we now derive. Consider the 
expectation value of fermionic vector current operator $j^{\mu}$
in the static $\sigx,~\pix$ background\footnote{In the following it is 
enough 
to discuss only the vector current, because the axial current 
$j_5^{\mu}=\epsilon^{\mu\nu}j_{\nu}$.}    

\beq 
\langle \sigx,~\pix| j^{\mu} | \sigx, ~\pix\rangle = -\int{d\om\over 
2\pi} {\rm tr}\left[~\gam^{\mu}~ \left(\begin{array}{cc} A(x) & 
B(x) \\{}&{}\\ C(x) & 
D(x)\end{array}\right)\right]\,.
\label{vectorcurrent}
\eeq
Therefore, we find from (\ref{asymptoticdirac}) that the asymptotic 
behavior 
of the current matrix elements is 
 
\beqra
\langle \sigx,~\pix| j^0 | \sigx, ~\pix\rangle &&\asymptotic 0
\nonumber\\
{\rm 
and}\quad\quad\quad\quad\quad\quad\quad\quad\quad\quad\quad\hfill
\nonumber\\
\langle \sigx,~\pix| j^1 | \sigx, ~\pix\rangle &&\asymptotic  
-\int{d\om\over 2\pi}
R\left(k\right)e^{2ik\,|x|}
\label{currents}
\eeqra
where we used the fact that $\int {d\om\over 2\pi} {\om\over k}
f(k) = 0$   because $k\left(\om\right)$ is an even function of $\om$.

Thus, an arbitrary static background $\sigx\,, \pix$ induces fermion 
currents that do not deacy as $x\rightarrow\pm\infty$, unless $R(k)\equiv 
0$.
Clearly, we cannot have such currents in our static problem and we 
conclude
that as far as the field theory (\ref{auxiliary}) is concerned, the 
fields
$\sigx\,, \pix$ must be such that the Sturm-Liouville operators in 
(\ref{bc}) and therefore the Dirac operator (\ref{dirac1}) are 
reflectionless.

The absence of reflections emerges here from basic principles of field 
theory, 
and not merely as a large $N$ saddle point condition, as in 
\cite{dhn,shei}.
Indeed, reflectionlessness of (\ref{dirac1}) must hold whatever the value 
of $N$ is. Therefore, the fact that reflectionlessness of (\ref{dirac1}) 
appeared in \cite{dhn,shei} as a saddle point condition in the inverse 
scattering formalism simply indicates consistency
of the large $N$ approximation in analysing space dependent condensations 
$\sigx\,, \pix$. The absence of reflections also 
restores asymptotic translational invariance. What we mean by this 
statement
is that if $R(k)\equiv 0$  then (\ref{asymptoticdirac}) is simply the 
result of inverting (\ref{dirac1}) in Fourier space with constant 
asymptotic background (\ref{boundaryconditions}), namely,

\beq
\langle x\,|-iD^{-1} | x\,\rangle = {1\over 2 \sqrt{m^2-\omega^2}}
\left(\begin{array}{cc} i m{\rm cos}\theta & \omega +m{\rm sin}\theta 
\\{}&{}\\ -\omega+m{\rm sin}\theta &  i m{\rm 
cos}\theta\end{array}\right)
\label{asymptotic}
\eeq
which therefore yields the asymptotic behavior of (\ref{diagonal}) for 
properly chosen chiral alignment angles. Note that in the absence of 
reflections, (\ref{asymptoticdirac}) attains its asymptotic value 
(\ref{asymptotic}) by simply following the asymptotic behavior of 
$\sigx$ and of $\pix$, which are the exclusive sources of any asymptotic
$x$ dependence of the resolvent. This expression (\ref{asymptotic}) has 
cuts in the complex $\omega$ plane stemming from scattering states of 
fermions of mass 
$m$. These cuts must obviously persist in $A, B, C$ and $D$ away from the 
asymptotic region, and we make use of this fact in the next section. 
We used the asymptotic matrix elements (\ref{vectorcurrent}) of the 
vector current operator in the background of static $\sigx,\,\pix$ to 
establish 
the absence of reflections in the static Dirac operator. We can now make 
use
of this result to examine its general dynamical implications on matrix
elements of other interesting operators, namely, the scalar 
$\bar\psi\psi$
and pseudoscalar $\bar\psi i\gam_5\psi$ density operators. Their matrix 
elements in the background of $\sigx,\,\pix$ 
are 
\beqra 
\langle \sigx,~\pix| \bar\psi\psi | \sigx, ~\pix\rangle &=& 
N\int{d\om\over 2\pi}~ {\rm tr}~\langle x\,|iD^{-1} | x\,\rangle 
\nonumber\\{\rm 
and}\quad\quad\quad\quad\quad\quad\quad\quad\quad\quad\quad\quad\quad\quad
\nonumber\\
\langle \sigx,~\pix| \bar\psi i\gam_5\psi | \sigx, ~\pix\rangle &=& 
N\int{d\om\over 2\pi}~ {\rm tr}\left[~i\gam_5~\langle x\,|iD^{-1} | 
x\,\rangle \right].
\label{scalarpseudoscalar}
\eeqra
Therefore, from (\ref{asymptoticdirac}) their asymptotic behavior is 
simply 
\beqra
\langle \sigx,~\pix| \bar\psi\psi | \sigx, ~\pix\rangle &\asymptotic & 
N\sigx~
\int{d\om\over 2\pi}~ {1 + R\left(k\right)e^{2ik\,|x|}\over k} 
\nonumber\\{\rm and}
\quad\quad\quad\quad\quad\quad\quad\quad\quad\quad\quad\quad\nonumber\\
\langle \sigx,~\pix| \bar\psi i\gam_5\psi | \sigx, ~N\pix\rangle 
&\asymptotic & \pix~\int{d\om\over 2\pi}~ {1 + 
R\left(k\right)e^{2ik\,|x|}\over k}\,.
\label{asymptoticdensity}
\eeqra
Clearly, in the absence of reflections, the asymptotic $x$ dependence
of these matrix elements follows the profiles of $\sigx$ and $\pix$, 
respectively. Otherwise, if $R(k)\neq 0 $, these matrix elements will
have further powerlike decay in $x$ superimposed on these profiles, which 
is 
not related directly to the typical length scales appearing in $\sigx$ 
and
in $\pix$. We close this section by investigating implications of 
(\ref{asymptoticdensity}) for extremal background configurations. For 
such 
configurations the matrix 
element of the scalar density is equal to $\sigx/g^2$ and that of the 
pseudoscalar density is equal to $\pix/g^2$. Such background fields 
must obviously correspond to a reflectionless Dirac operator, but let us 
for the moment entertain ourselves with the assumption that $R(k)$ in 
(\ref{asymptoticdensity}) is arbitrary and see how the absence of
reflections appears as a saddle point condition. Thus, for extremal 
configurations, as $x\rightarrow\pm\infty$, $\sigx$ cancels off both 
sides of 
the first equation in (\ref{asymptoticdensity}) and $\pix$ cancels off 
both 
sides of the other equation. This leaves us with a common dispersion 
integral

\beqast
{1\over Ng^2} = \int{d\om\over 2\pi}~ {1 + 
R\left(k\right)e^{2ik\,|x|}\over k}\,.
\eeqast
It turns out (see (\ref{spectral} below) that the integral over the first 
$x$ independent term on the right hand side cancels precisely the
constant term on the left hand side. This is simply a reformulation of 
(\ref{bgap}) in Minkowski space. Therefore, the remaining $x$ dependent
integral must vanish for any (large) $|x|$. It follows then, that $R(k)$ 
must vanish. Thus absence of reflections appears here as a saddle point 
requirement, in a rather simple elegant manner, without ever invoking the 
inverse scattering transform. The whole purpose of this section is to 
prove that one cannot consider static reflectionful backgrounds to begin 
with, and
thus the emergence of absence of reflections as a saddle point
condition is simply a successful consistency check for the validity of 
the
large $N$ approximation applied to space dependent condensates.

\pagebreak

\section{The Diagonal Resolvent for a Fixed Number of Bound States}
\setcounter{equation}{0}

The requirement that the static Dirac operator (\ref{dirac1}) be 
reflectionless is by itself quite restrictive, since most $\sigx,\pix$ 
configurations will not lead to a 
reflectionless static Dirac operator. Construction of explicit 
expressions for the resolvent 
in terms of $\sigx\,,\pix$ and their derivatives is a formidable task
even under such severe restrictions on these fields. We now show how to 
accomplish such a construction at the price 
of posing further restrictions on $\sigx$ and $\pix$ in function space.
However, even under these further restrictions the results we obtain are 
still quite interesting from a physical point of view.

In the following we concentrate on the $B(x)$ component of 
(\ref{diagonal}).
The other entries in (\ref{diagonal}) may be deduced from $B(x)$ trough
(\ref{abcd}) and (\ref{bcrelation}). 
 
Our starting point here is the observation that one can derive from the 
representation of $B(x)$ in (\ref{abcd}) a functional identity in the 
form of a differential equation relating $B(x)$ to $\sigx$ and $\pix$ 
without ever knowing the 
explicit form of the Jost functions $b_1(x)$ and $b_2(x)$. We leave the 
details of derivation to Appendix B, where we show that the identity 
mentioned above is 

\beqra
&&\pax\left\{{1\over \omega + \pix} \pax\left[{\pax B(x)\over 
\omega+\pix}\right]\right\}\nonumber\\{}\nonumber\\ 
&-& {4\over \omega + \pix} \left\{\pax\left[{B(x)\over 
\omega+\pix}\right]\right\}\left[\sigx^2+\pix^2-\si'(x)-\omega^2+
{\sigx\pi'(x)\over\omega+\pix}\right]\nonumber\\{}\nonumber\\ &-& 
{2B(x)\over 
\left[\omega+\pix\right]^2 }\, \pax\left[\sigx^2+\pix^2-\si'(x) 
+ {\sigx\pi'(x)\over \omega+\pix}\right]\,\equiv\,0
\label{gdblinear}
\eeqra
with a similar expression for $C(x)$ in which 
$\si\rightarrow -\si\,\quad \pi\rightarrow -\pi$ 
that we do not write down explicitly. 

Here we denote derivatives with respect to $x$ either by primes or by 
partial derivatives. This equation is a linear form of what is referred 
to in the mathematical literature as the ``Gel'fand-Dikii" 
identity\cite{gd}.  This identity merely reflects the fact that $B(x)$ is 
the diagonal resolvent of the Strum-Liouville operator discussed above 
and sets no restrictions on $\sigx$ and $\pix$. 

If we were able to solve (\ref{gdblinear}) for $B(x)$ in a closed form 
for any static configuration of $\sigx, \pix$, we would then be able to 
express $\langle x\,|iD^{-1} | x\,\rangle $ in terms of the latter
fields and their derivatives, and therefore to integrate (\ref{saddle})
back to find an expression for the effective action (\ref{effective}) 
explicitly in terms of $\sigx$ and $\pix$. Invoking at that point Lorentz 
invariance of (\ref{effective}) we would then actually be able to write 
down the full effective action for space-time dependent $\si$ and $\pi$. 
Note moreover that in principle such a procedure would yield an exact 
expression for the effective action, regardless of what $N$ is. 

Unfortunately, deriving such an expression for $B(x)$ is a difficult
task, and thus we set ourselves a simpler goal in this paper, by 
determining the
desired expression for $B(x)$ with $\sigx, \pix$ restricted to a specific 
sectors in the 
space of all possible static configurations. 
To specify these sectors consider the Dirac equation 
associated with (\ref{dirac}), $D\psi=0$. For a given configuration of 
$\sigx, \pix$ (such that $D$ is reflectionless), this equation has $n$ 
bound states at energies $\omega_1, 
..., \omega_n$ as well as scattering states. A given sector is then 
defined by specifying the number of bound states the Dirac equation 
has.\footnote {The effect of scattering
states on $B(x)$ is rigidly fixed by spatial asymptotics, as 
(\ref{asymptotic})
indicates, so only bound states are used to specify such a sector.}

As we saw above, $B(x)$ must have a cut in the $\omega$ plane with branch 
points at $\omega=\pm m$. If in addition to scattering states $\sigx, 
\pix$ support 
$n$ bound states at energies $\omega_1, ... \omega_n$ (which must all lie 
in the real interval $-m<\omega<m$)\footnote{The Gross-Neveu 
model\cite{gn,dhn,josh1} is a theory of Majorana (real) fermions. 
Therefore its spectrum is invariant under charge conjugation, i.e., it is 
symmetric under $\omega\rightarrow -\omega$. Thus in that case the bound 
states are paired symmetrically around $\omega=0$ and $B(x)$ is really a 
function of $\omega^2$. The chiral NJL model on the other hand is a 
theory of Dirac (complex) fermions, charge conjugation symmetry of the 
spectrum is  broken by the $\pi$ field and bound states are not paired.} 
then the 
corresponding $B$ must contain a simple pole for each of these bound 
states. Therefore, $B(x)$ must contain the purely $\omega$ dependent 
factor 
\beq
{1\over  \sqrt{m^2-\omega^2} \prod_{k=1}^n (\omega-\omega_k)}
\label{factor}
\eeq
of mass dimension $-n-1$. Any other singularity $B(x)$ may have in the 
complex $\omega$ plane cannot be directly related to the spectrum of the 

Dirac operator, and therefore must involve $x$ dependence as well. Based 
on 
our discussion in Appendix A, the only possible combination that mixes 
these variables is ${\rm exp}(i\sqrt{\om^2-m^2}~ x)$. But such a 
combination is ruled out as we elaborated in the previous section, by the 
requirement that the Dirac operator be reflectionless.
The factor (\ref{factor}) then exhausts all allowed singularities of 
$B(x)$ 
in the complex $\om$ plane. Recall further that $B(x)$ is a dimensionless 
quantity, and thus the negative dimension of the $\omega$ dependent 
factor (\ref{factor}) must be balanced by a polynomial of degree $n+1$ in 
$\omega$ (with $x$ dependent coefficients) of 
mass dimension $n+1$, namely\footnote{One may argue that Eq. (\ref{bofx}) 
should be further multiplied by a dimensionless bounded function 
$f({\omega\over m})$. However such a function must be entire, otherwise 
it will changed the prescribed singularity properties of $B(x)$, but the 
only bounded entire functions are constant.}
\beq
B(x,\om) = { {B_{n+1}(x)\omega^{n+1} + .... +  B_{1}(x)\omega +  
B_{0}(x)}\over
  {\sqrt{m^2-\omega^2} \prod_{k=1}^n (\omega-\omega_k)}} \,. 
\label{bofx}
\eeq
The mass dimension of $B_k (x) ~(k=0,...,n+1) $  is $n+1-k$. 

The main point
here is that simple dimensional analysis in conjunction with the 
prescribed analytic properties of $B(x)$ fix its $\omega$ dependence 
completely, up to $n+1$ unknown bound state energies, and $n+2$ unknown 
functions of 
$\sigx, \pix$ and their derivatives. These functions are by no means 
arbitrary. They have to be such that (\ref{bofx}) and the resulting 
expression for 
$C(x)$  are indeed the resolvents of the appropriate Sturm-Liouville  
operators. These expressions for $B(x)$ and $C(x)$ must be therefore 
subjected to the 
Gel'fand-Dikii identities (\ref{gdblinear}) and the corresponding 
identity 
for
$C(x)$.   

Substituting $B(x)$ into (\ref{gdblinear}) we obtain an equation of the
form 
\beq Q^{\left(B\right)}_{n+5}\left(\omega,x\right)/
\left[\omega+\pix\right]^4 \equiv 0\,, \label{qeq} 
\eeq 
where
$Q^{\left(B\right)}_{n+5}\left(\omega,x\right)$ is a polynomial of degree
$n+5$ in $\omega$ with $x$ dependent coefficients that are linear
combinations of the functions $B_k(x)$ and their first three derivatives.

Note that because of the linearity and homogeneity of (\ref{gdblinear}),
the purely $\omega$ dependent denominator of (\ref{bofx}) with its
explicit dependence on the bound state energies drops out from
(\ref{qeq}). This is actually the main advantage\footnote{The non-linear
version of the Gel'fand-Dikii identity (\ref{slgd}) (or (\ref{gdb}))
contains further information about the normalisation of $B(x)$, but the
latter may be readily determined from the asymptotic behavior
(\ref{asymptotic}) of $B(x)$.} of working with the linear form of the
Gel'fand-Dikii identity rather than with its non-linear form
(\ref{gdb})\cite{josh1}.

Setting to zero each of the $x$ dependent coefficients in
$Q^{\left(B\right)}_{n+5}$ we obtain an overdetermined system of
$n+6$ linear differential equations in the $n+2$ functions $B_k(x)$.
Using $n+2$ of the equations we fix all the functions $B_k(x)$ in terms 
of
$\sigx$, $\pix$ and their derivatives, up to $n+2$ integration constants
$b_k$. These integration constants are completely determined once we
enforce on the resulting expression for $B(x)$ the asymptotic behavior
(\ref{boundaryconditions}) and (\ref{asymptotic}). The integration
constants $b_k$ turn out to be polynomials in $m^2$ and the bound state
energies $\omega_k$.

At this stage we are left, independently of $n$,  with four non-linear 
differential equations in $\sigx$ and $\pix$\footnote{Coefficients of the 
various terms in these equations are also polynomials in $m^2$ and the 
bound state energies $\omega_k$.}.  
A similar analysis applies for $C(x)$, leading to an equation of the form
$Q^{\left(C\right)}_{n+5}\left(\omega,x\right)\equiv\,0$, 
where following (\ref{bcrelation}) 
$Q^{\left(C\right)}_{n+5} \left(\omega,\sigx,
\pix\right)= - Q^{\left(B\right)}_{n+5} \left(\omega,-\sigx,
-\pix\right)$\,. 
Setting the first $n+2$ coefficients in
$Q^{\left(C\right)}_{n+5}\left(\omega,x\right)$ to zero we verify that
$C(x)$ is related to $B(x)$ as in (\ref{bcrelation}), but that the
remaining four equations for $\sigx$ and $\pix$ are different from their
counterparts associated with $B(x)$ as the explicit relation between
$Q^{\left(C\right)}_{n+5}$ and $Q^{\left(B\right)}_{n+5}$ suggests. 
We are thus left with an overdetermined set of eight non-linear 
differential
equations for the two functions $\sigx$ and $\pix$. Observing that
$Q^{\left(C\right)}_{n+5}\pm Q^{\left(B\right)}_{n+5}$ is odd (even) in
$\si$ and $\pi$, we note that these eight equations are equivalent to
breaking each of the four remaining equations associated with $B(x)$ into
a part even in $\si$ and $\pi$ and a part odd in $\si$ and $\pi$ and
setting each of these parts to zero separately. 

Mathematical consistency of our analysis requires that the six most 
complicated equations of the total eight be redundant relative to the 
remaining two equations, because we have only two unknown functions, 
$\sigx$ and $\pix$. This requirement must be fulfilled, because 
otherwise we are compelled to 
deduce that there can be no $\sigx$ and $\pix$ configurations for which 
the Dirac equation $D\psi=0$ has precisely $n$ bound states, with 
$n=0, 1, 2, \cdots$, which is presumably an erroneous conclusion.
 
Therefore $\sigx$ and $\pix$ are uniquely determined from the two 
independent equations given the asymptotic boundary condittions
(\ref{boundaryconditions}) they satisfy. This leaves only the bound state 
energies undetermined, but the latter cannot be determined by the 
resolvent identity, which does not really care what their values are. 
These energies are determined by imposing the saddle point conditions 
(\ref{saddle}), i.e., by  dynamical aspects of the model under 
investigation.

In the preceding paragraphs we laid down the mathematical aspects of our 
analysis. We now add to these a symmetry argument which will 
simplify our solution of the differential equations for $\sigx$ and 
$\pix$ 
a great deal. The two non-redundant coupled differential equations for 
$\sigx$ 
and $\pix$ allow us to eliminate one of these functions in terms of the 
other. 
We choose to eliminate\footnote{We prefer to eliminate $\pix$ in terms of 
$\sigx$ because the latter never vanishes identically.} $\pix$ in terms 
of 
$\sigx$,
\beq
\pi_{\alpha}(x) = G_{\alpha}[\si_{\alpha}(x)]
\label{pisigma1}
\eeq
where $\alpha$ is a global chiral alignment angle. This relation is 
clearly covariant
under axial rotations $\alpha\rightarrow \alpha + \Delta\alpha$, because 
$\sigx$ and $\pix$ transform as the two components of a vector under 
$U(1)_A$ as 
(\ref{vector}) shows. We expect (\ref{pisigma1}) to be a 
 linear relation. Imposing the boundary conditions 
(\ref{boundaryconditions}) we have
\beq
\pix=-[\sigx-m]\,{\rm cot}{\theta\over 2}\,.
\label{pisigma}
\eeq

In this way we reduce the problem into finding the single function 
$\sigx$. 
The condition (\ref{pisigma}) is an external supplement to the coupled
differential equations for $\sigx$ and $\pix$ stemming from the 
Gel'fand-Dikii equation. We thus have to make sure that the resulting 
solution for $\sigx$ 
and (\ref{pisigma}) are indeed solutions of these coupled differential 
equations.

We now provide the details of such calculations in the case of a single 
bound state, as well as partial results concerning two bound states.

\pagebreak

\section{ Extended Object Profiles}
\setcounter{equation}{0} 
\subsection{ A single bound state}

In this case (\ref{bofx}) becomes 
\beq
B(x) = { {B_2 (x)\omega^2  +  B_1 (x)\omega +  B_0 (x)}\over
{\left(\omega-\omega_1\right)\sqrt{m^2-\omega^2}}}  
\label{bofx1}
\eeq
where the single bound state energy is $\omega_1$. Then setting to zero 
the 
coefficients of $\omega^6$ through $\omega^4$ in the degree six 
polynomial 
(\ref{qeq}), we find
\beqra
B_2 (x)&=&b_2\,,\quad\quad B_1 (x)=b_2 \pix + b_1 \quad\quad {\rm 
and}\nonumber\\{}\nonumber\\
B_0 (x)&=& b_1 \pix + {b_2 \over 2}~ [\si^2(x) + 
\pi^2(x) - \si^{'} (x)] + b_0
\label{bofx11}
\eeqra  
where $b_2, b_1$ and $b_0$ are integration constants. We then impose the 
asymptotic boundary conditions 
\beqast
B(x)\asymptotic  {1\over 2 \sqrt{m^2-\omega^2}}
\left(\omega +m{\rm sin}\theta_{\pm}\right)
\eeqast
to fix the latter,  
\beq
b_2 = {1\over 2}\,,\quad\quad b_1 = -{\om_1 \over 2}\,,\quad\quad b_0 = 
-{m^2 \over 4}
\label{b2b1b0}
\eeq 
and therefore
\beq
B(x) = { \omega  +  \pix \over
2 \sqrt{m^2-\omega^2}} +   {\si^2(x) + \pi^2(x) - \si^{'} (x) - m^2 
\over
4\left(\omega-\omega_1\right) \sqrt{m^2-\omega^2}}\,.
\label{bofx1final}
\eeq

The relation (\ref{bcrelation}) then immediately leads to 
\beq
C(x) = -{ \omega  -  \pix \over
2 \sqrt{m^2-\omega^2}} -   {\si^2(x) + \pi^2(x) + \si^{'} (x) - m^2 
\over
4\left(\omega-\omega_1\right) \sqrt{m^2-\omega^2}}\,.
\label{cofx1final}
\eeq

Having the coefficients of $\om^6$ through $\om^4$ in the degree six 
polynomial (\ref{qeq}) set to zero, we are left with a cubic polynomial 

\beqra
4\pax\left\{\left[\left(m^2-\pi^2(x)-\si^2(x)\right)\pix-\om_1\si{'}(x)+{
1\over 
2}\pi^{''}(x)\right]\right.+\nonumber\\
\left.\left[\om_1\left(\pi^2(x)+\si^2(x)\right)-\sigx\pi^{'}(x)+
\si^{'}(x)\pix\right]\right\}\om^3+\cdots\,\equiv\,0
\label{cubic}
\eeqra
where the $(\cdots)$ stand for lower powers of $\om$. The cubic 
(\ref{cubic}) has to be set to zero identically, producing eight coupled 
differential equations in $\sigx$ and $\pix$ as we discussed above. The 
simplest equation of these is obtained by setting to zero the part of the 
$\om^3$ coefficient in 
(\ref{cubic}) that is even in $\si$ and $\pi$, namely,
\beqast
\pax\left[\om_1\left(\pi^2(x)+\si^2(x)\right)-\sigx\pi^{'}(x)+
\si^{'}(x)\pix\right]\,=\,0
\eeqast
which we immediately integrate into
\beq
\om_1\left[\pi^2(x)+\si^2(x)-m^2\right]\,=\,\sigx\pi^{'}(x)-\si^{'}(x)\pi
(x)\,.
\label{simplest1}
\eeq

Here we have used the boundary conditions (\ref{boundaryconditions}) to 
determine the integration constant. 
The next simplest equation is obtained by setting to zero the part of the 
$\om^3$ coefficient in (\ref{cubic}) that is odd in $\si$ and $\pi$, and 
so on. 

Following our general discussion we solve the system of coupled equations
(\ref{pisigma}) and (\ref{simplest1}), which leads to 
\beq
{d\over dx}\left[{1\over \sigx-m}\right] - {2\om_1 \,{\rm 
tan}{\theta\over 
2} \over \sigx - m}\,=\, {2\om_1\over m {\rm sin} \theta}\,.
\label{sieq}
\eeq

Solving (\ref{sieq}) we find
\beqra
\sigx &=& m - {m\,{\rm sin} \theta \,{\rm tan} {\theta\over 2} \over 1 + 
{\rm exp}\left[2\om_1 {\rm tan} {\theta\over 2}\cdot 
\left(x-x_0\right)\right]}\nonumber\\{}\nonumber\\
\pix &=& {m {\rm sin} \theta \over   1 + {\rm exp}\left[2\om_1 {\rm tan} 
{\theta\over 2} \cdot\left(x-x_0\right)\right]}
\label{sigmapi1bs}
\eeqra
where we have chosen the integration constant (parametrised by $x_0$) 
such 
that $\sigx$ and $\pix$ 
would be free of poles. Note that the boundary conditions at 
$x\rightarrow 
+\infty$ require
\beq
\om_1 \,{\rm tan} {\theta\over 2} < 0\,.
\label{negative}
\eeq

Substituting the expressions (\ref{sigmapi1bs}) into (\ref{bofx1final}) 
one finds  that the resulting $B(x)$ is indeed a solution of the 
corresponding 
Gel'fand-Dikii equation (\ref{gdblinear}), verifying the consistency of 
our solution. Our results (\ref{sigmapi1bs}) for $\sigx$ and $\pix$ agree 
with those of \cite{shei}. They have the profile of an extended object, a 
lump or a chiral ``bag",  of size of the order ${\rm cot} {\theta\over 
2}/\om_1$ 
centred around an arbitrary point $x_0$. Note that the profiles in 
(\ref{sigmapi1bs})
satisfy 
\beq
\rho^2(x) = \si^2(x) + \pi^2(x) = m^2 - m^2\,{\rm 
sin}^2\,(\theta/2)\,{\rm sech}^2\,\left[\om_1 {\rm tan} 
{\theta\over 2} \cdot\left(x-x_0\right)\right]\,.
\label{rho1bs}
\eeq
Thus, as expected by construction, this configuration interpolates 
between 
two different vacua at $x=\pm\infty$. As $x$ increases from $-\infty$, 
the vacuum configuration becomes distorted. The distortion reaches its 
maximum at the location of the ``bag", where $m^2-\rho^2(x_0) =  
m^2\,{\rm sin}^2\,(\theta/2)$ and then relaxes back into the other vacuum 
state at $x=\infty$.  The arbitrariness of $x_0$ is, of course, a 
manifestation of translational invariance. 
\pagebreak

\subsection{Two Bound States}
In this case (\ref{bofx}) becomes 
\beq
B(x) = { {B_3 (x)\omega^3 +  B_2 (x)\omega^2  +  B_1 (x)\omega +  B_0 
(x)}\over
{\left(\omega-\omega_1\right)\left(\omega-\omega_2\right)\sqrt{m^2-\omega
^2}}}  
\label{bofx2}
\eeq
where the bound state energies are $\omega_1$ and $\om_2$. (Obviously, 
$B(x)$ in this subsection should not be confused with its counterpart in 
the previous subsection.) In this case 
the polynomial (\ref{qeq}) is of degree seven in $\om$. Following the 
procedure outlined in Section 3 we find after imposing the boundary 
conditions
(\ref{boundaryconditions}) that  
\beqra
&&B(x) = { \omega  +  \pix \over
2 \sqrt{m^2-\omega^2}} + \hfill 
\nonumber\\{}\nonumber\\
&&{{[\si^2(x) + \pi^2(x) - \si^{'} (x) - m^2] [\pix + \om - \om_1 - 
\om_2]}\over
4\left(\omega-\omega_1\right) 
\left(\omega-\omega_2\right)\sqrt{m^2-\omega^2}}
\nonumber\\{}\nonumber\\
&&-{\pi''\left(x\right)-2\sigx\pi'\left(x\right)\over
8\left(\omega-\omega_1\right) 
\left(\omega-\omega_2\right)\sqrt{m^2-\omega^2}}\,.
\label{bofx2final}
\eeqra

Then, from  (\ref{bcrelation}) we find that  
\beqra
&&C(x) = - { \omega  -  \pix \over
2 \sqrt{m^2-\omega^2}} - \hfill 
\nonumber\\{}\nonumber\\
&&{{[\si^2(x) + \pi^2(x) + \si^{'} (x) - m^2] [-\pix + \om - \om_1 - 
\om_2]}\over
4\left(\omega-\omega_1\right) 
\left(\omega-\omega_2\right)\sqrt{m^2-\omega^2}}
\nonumber\\{}\nonumber\\
&&-{\pi''\left(x\right)+2\sigx\pi'\left(x\right)\over
8\left(\omega-\omega_1\right) 
\left(\omega-\omega_2\right)\sqrt{m^2-\omega^2}}\,.
\label{cofx2final}
\eeqra

Note that if we set $\om_1+\om_2=0$ and $\pix = 0$ the resolvents 
(\ref{bofx2final}) and (\ref{cofx2final}),  and therefore the whole 
spectrum, become invariant under $\om\rightarrow -\om$, and we obtain the 
equation appropriate to the Gross-Neveu model.

Setting to zero the coefficients of $\om^7$ through $\om^4$ in the degree 
seven 
polynomial (\ref{qeq}), we are left with a cubic polynomial in 
$\om$ which we do not write down explicitly. This polynomial must vanish 
identically, producing eight coupled 
differential equations in $\sigx$ and $\pix$ as we discussed above. The 
simplest equation of these is obtained by setting to zero the part of the 
$\om^3$ coefficient in that polynomial which unlike the previous case, 
is now odd in $\si$ and $\pi$
\beqra
&&2\pax\left\{\,4\,(\om_1+\om_2)\,[\pi^2(x)+\si^2(x)-m^2]\,\pix 
+\,6\,[\pi^2(x)+\si^2(x)]\,\si'(x)\right.
\nonumber\\
&&\left.+2\,(2\om_1\om_2-m^2)\,\si'(x) - 2\,(\om_1+\om_2)\,\pi^{''}(x)-
\si^{'''}(x)\right\}\,=\,0\,.
\eeqra
As in the previous case, this is a complete derivative which we readily 
integrate into
\beqra
&&4\,(\om_1+\om_2)\,[\pi^2(x)+\si^2(x)-m^2]\,\pix 
+\,6\,[\pi^2(x)+\si^2(x)]\,\si'(x)
\nonumber\\
&&+2\,(2\om_1\om_2-m^2)\,\si'(x) - 2\,(\om_1+\om_2)\,\pi^{''}(x)-
\si^{'''}(x)\,=\,0\,.
\label{simplest2}
\eeqra
Here we have used the boundary conditions (\ref{boundaryconditions}) to 
determine the integration constant. Note that (\ref{simplest2}) is of 
third
order in derivatives and cubic, whereas its single bound state 
counterpart (\ref{simplest1})
is only first order in derivatives and quadratic.

The next simplest equation is obtained by setting to zero the part of the 
$\om^3$ coefficient in the cubic polynomial that is evem in $\si$ and 
$\pi$, 
\beqra
&&2\,\pax\{\,2\,(m^2-2\om_1\om_2)\,[\pi^2(x)+\si^2(x)] 
-\,3\,[\pi^2(x)+\si^2(x)]^2 + 4(\om_1+\om_2)
\nonumber\\
&&\cdot [\sigx\pi'(x)-\si'(x)\pix]\}
+4\,[\pix\pi^{'''}(x) + \sigx\si^{'''}(x)]\,=\,0\,.
\eeqra
and so on. 

Following our general discussion we have to solve solve the system of 
coupled equations
(\ref{pisigma}) and (\ref{simplest2}) which is equivalent to 
\beqra
&&2\,\la\,(\om_1+\om_2)\,[4m\,y^2 + 2(1+\la^2) y^3 - y'']
\nonumber\\
&&+\pax\{\,4\,(m^2+\om_1\om_2)\,y + 6\,m\,y^2 + 2\,(\la^2+1)\,y^3 - y''\} 
= 0  
\label{sieq2}
\eeqra
where $\la = -{\rm cot} (\theta/2)$ and $y(x)=\sigx-m$. We have not 
succeeded
in solving this non-linear ordinary differential equation in closed form.

\pagebreak

\section{The Saddle Point Conditions}

Derivation of the explicit expressions of $\sigx$ and $\pix$ does not 
involve the saddle point equations (\ref{saddle}). 
Rather, it tells us independently of the large $N$ approximation that 
$\sigx$ and $\pix$ must have the form given in 
(\ref{sigmapi1bs}) in order for the associated Dirac 
operator to be reflectionless and to have a single bound state at a 
prescribed energy $\om_1$ in addition to scattering states. Thus, for the 
solution (\ref{sigmapi1bs}) we have yet to determine the values of 
$\om_1$
and $\theta$ allowed by the saddle point condition. More generallly, 
our discussion in Section III will lead us to the $\sigx$ and $\pix$ 
configurations which correspond to reflectionless Dirac operators with a 
prescribed number of bound states at some prescribed energies $\om_1, 
\om_2, \cdots$ in addition to scattering states. As emphasized earlier, 
this 
result is independent of the large $N$ limit. The allowed values of 
$\theta, \omega_1, \om_2, \cdots$ must then be determined by the saddle 
point condition (\ref{saddle}). It is this dynamical feature that we can 
analyse 
only in the large $N$ limit. 

For static background fields the general saddle point condition 
(\ref{saddle}) assume the simpler form
\begin{eqnarray}
\si(x) + ~Ng^2 \int\,{d\omega\over 2\pi} \left[A(x)+D(x)\right] &=& 0
\nonumber\\{}\nonumber\\
\pi(x) + iNg^2 \int\,{d\omega\over 2\pi} \left[B(x)+C(x)\right]\ &=& 0\,.
\label{saddle1}
\end{eqnarray}
In Subsection 5.1 we impose this condition on the explicit single bound 
state background we found in the previous section and calculate the mass 
of such ``bags". The two bound states case is discussed in Subsection 
5.2.

\subsection{A single bound state}

Substituting (\ref{bofx1final}), (\ref{cofx1final}) into the saddle point 
equations (\ref{saddle1}) we obtain\footnote{In the following formula we 
omit explicit $x$ dependence of the fields. The number $\pi$ 
also appears in the formula, but only in the combination ${d\om\over 
2\pi}$. Therefore there is no danger of confusing  the field $\pix$  and 
the number $\pi$.} 

\begin{eqnarray}
&&{\si\over Ng^2} + ~i\int\limits^{\Lambda}_{-\Lambda}\,{d\omega\over 
2\pi}{1\over 4\sqrt{m^2-\om^2} 
\left(\om-\om_1\right)\left(\om^2-\pi^2\right)}
\left\{4\si\omega^3 + 
2\left(\pi'-2\om_1\si\right)\om^2  \right.
\nonumber\\{}\nonumber\\ 
&&\left. + 2\left[\si\left(\si^2-\pi^2-m^2\right)-\om_1\pi' -{1\over 
2}\si''\right]\om - 2\pi^2\left(\pi'-2\om_1\si\right) \right\} = 0 
\nonumber\\{}\nonumber\\
&&{\pi \over Ng^2} + i\int\limits^{\Lambda}_{-\Lambda}\,{d\omega\over 
2\pi} 
{2\pi\om-2\om_1\pi-\si' \over 2\sqrt{m^2-\om^2}\left(\om-\om_1\right)} = 
0\,.
\label{saddle1bs}
\end{eqnarray}

These equations are dispersion relations among the various $x$ dependent 
parts and $\om_1$. Clearly, both dispersion integrals in 
(\ref{saddle1bs}) 
are logarithmically divergent in $\Lambda$, but  subtracting each of the 
integrals once we can get rid of these divergences. The required 
subtractions are actually already built in in (\ref{saddle1bs}). In order 
to see this consider the (bare) gap equation (\ref{bgap}) in Minkowski 
space. This equation is equivalent to 
the (logarithmically divergent) dispersion relation\footnote{To see this 
equivalence simply perform (contour) integration over spatial momentum 
first.}
 
\beq
{i\over Ng^2} = \int\limits^{\Lambda}_{-\Lambda}\, {d\om \over 2\pi}
{1\over \sqrt{m^2-\om^2 +i\epsilon}}\,.
\label{spectral}
\eeq

If we now replace each of the 
${1\over Ng^2}$ coefficients in (\ref{saddle1bs}) by the integral on the 
right hand 
side of (\ref{spectral}) \cite{josh1}, the divergent parts of 
each pair of integrals cancel and the equations (\ref{saddle1bs}) become
\begin{eqnarray}
\int\limits_{\cal C}\,{d\omega\over 2\pi}~ {\pi' 
+{\om\over\om^2-\pi^2}F(\si, \pi) \over  \sqrt{m^2-\om^2} 
\left(\om-\om_1\right)} &=& 0\nonumber\\{}\nonumber\\
\int\limits_{\cal C}\,{d\omega\over 2\pi}~ {\si' \over  \sqrt{m^2-\om^2} 
\left(\om-\om_1\right)} &=& 0
\label{saddlepair}
\end{eqnarray}
where 
\beq 
F(\si, \pi) = \si(\si^2+\pi^2-m^2)-\om_1\pi'-{1\over 2}\si''
\label{fox}
\eeq  
and ${\cal C}$ is the contour in the complex $\omega$ plane depicted in 
Fig.(1).
\vspace{24pt}
\par
\hspace{0.5in} \epsfbox{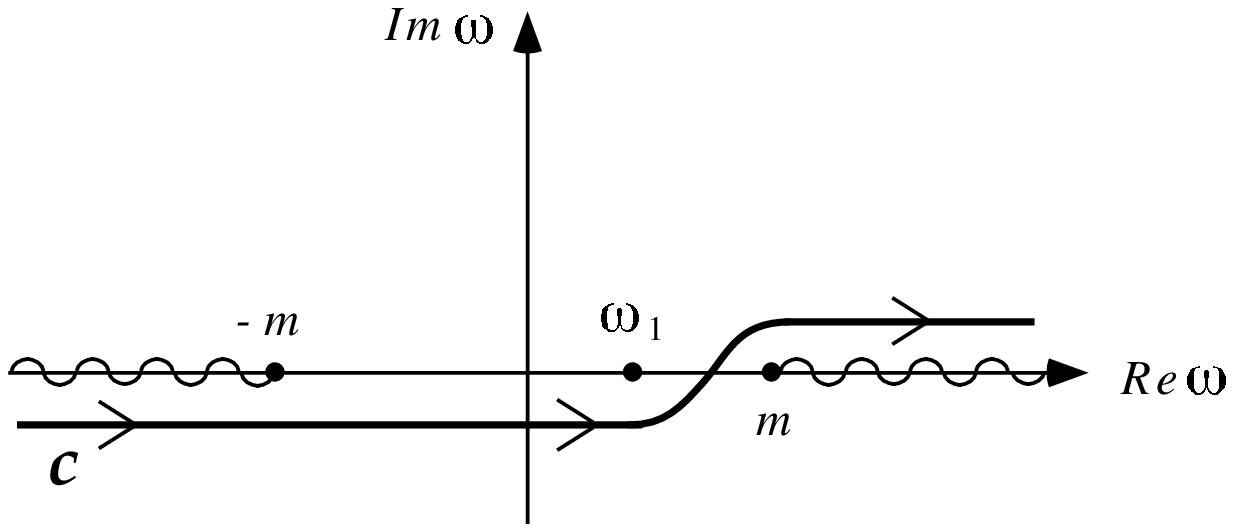}
\par
\baselineskip 12pt
{\footnotesize {\bf Fig.~1:} The contour ${\cal C}$ in the complex $\om$
plane in Eq.
(\ref{saddlepair}). The continuum states appear as the two cuts along the
real axis with branch points at $\pm m$ (wiggly lines), and the bound
state is the pole at $\om$.}
\vspace{24pt}
\par
\baselineskip 20pt   
The expression (\ref{fox}) is the residue of the $x$ dependent poles at 
$\om = \pm\pix$ in the first equation in (\ref{saddlepair}). The 
quantisation condition (\ref{saddlepair}) on $\om_1$ cannot be $x$ 
dependent. Therefore 
(\ref{fox}) must vanish as a consistency requirement. 
Substituting\footnote{Here we have set $x_0=0$ for simplicity.}  
(\ref{sigmapi1bs}) in (\ref{fox}) we find that 
\beqast
F(x) = {m\,{\rm tan}^2\, {\theta \over 2}\over 2}\,\, {\rm sech}^3 
\left[\om_1\,x{\rm tan} {\theta \over 2}\right]\left[{\rm cos}\theta 
e^{-\om_1\,x{\rm tan} {\theta \over 2}} + e^{\om_1\,x{\rm tan} {\theta 
\over 2}}\right]\left(\om_1^2-m^2{\rm cos}^2{\theta \over 2}\right)\,.
\eeqast 
Thus, $F(\si, \pi)$ vanishes for the 
configurations (\ref{sigmapi1bs}) provided 

\beq
\om_1^2 = m^2 {\rm cos}^2 ({\theta\over 2})
\label{luckily}
\eeq
which sets an interesting relation between the bound state energy and the 
chiral alignment angle of the vacuum at $x\rightarrow +\infty$. This 
relation
actually leaves $\theta$ the only free parameter in the problem with 
respect to which we have to extremise the action. The condition 
(\ref{negative}) then 
picks out one branch of (\ref{luckily}). 

We still have to determine $\om_1$ and $\theta$ separately. Now 
the saddle point condition simply boils down to the 
single equation
\beq
I(\om_1)=\int\limits_{\cal C}\,{d\omega\over 2\pi} {1 \over  
\sqrt{m^2-\om^2} 
\left(\om-\om_1\right)}\,=\,0\,.
\label{saddle1bsfinal}
\eeq

The contour integral in (\ref{saddle1bsfinal}) is most conveniently 
calculated
by deforming the contour ${\cal C}$ into the contour ${\cal C}'$ shown in 
Fig.(2). The ``hairpin" wing of ${\cal C}'$ picks up the contribution of 
the filled Fermi sea, and the little circle arount the simple pole at 
$\om=\om_1$ 
is the contribution of fermions populating the bound state of the ``bag".
\par
\vspace{24pt}
\hspace{0.5in} \epsfbox{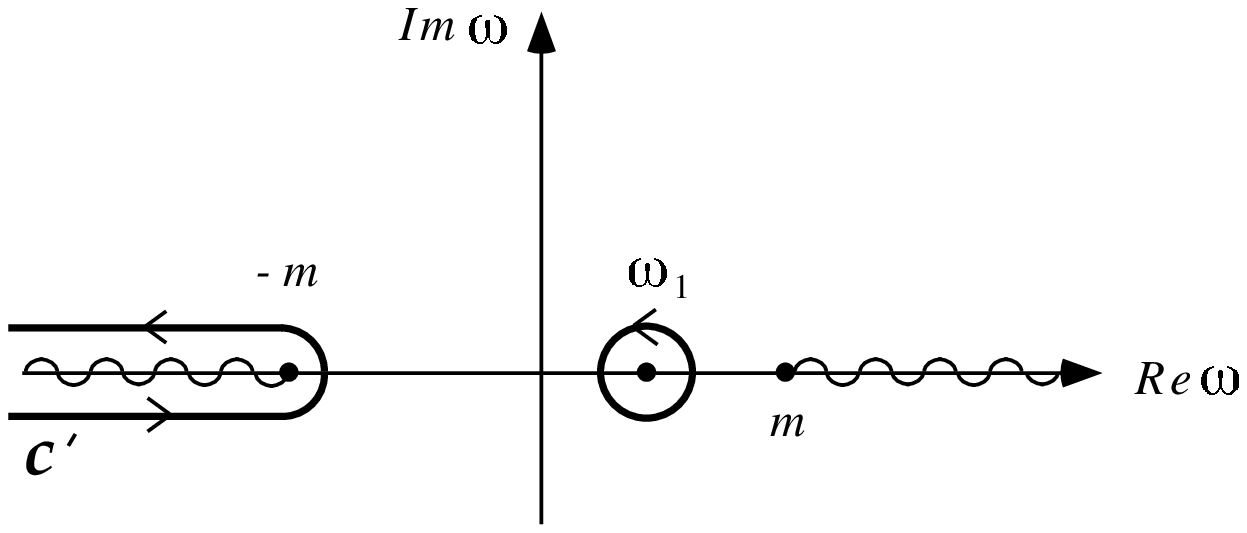}
\par
\baselineskip 12pt
{\footnotesize {\bf Fig.~2:} The deformed integration contour ${\cal C}'$
leading to Eq. (\ref{arctan}).}
\vspace{24pt}
\par
\baselineskip 20pt

Assuming that the ``bag" traps $n$ fermions in that state, and recalling 
that each state in the Fermi sea continuum accomodates $N$ fermions, we 
see that (\ref{saddle1bsfinal}) becomes
\beq
I(\om_1)={iN\over\sqrt{m^2-\om_1^2}}\left[{2\over\pi}\,{\rm 
arctan}\sqrt{{m+\om_1\over m-\om_1}} + {n\over N} - 1\right]\,=\,0
\label{arctan}
\eeq

The solution of this equation yields the quantisation condition
\beq
\om_1 = m {\rm cos}\left(n\pi\over N\right)
\label{quantisation1bs}
\eeq
in agreement with \cite{shei}. 

It follows from (\ref{sigmapi1bs}), (\ref{negative}) and (\ref{luckily}) 
that 
\beq
\theta = - {2\pi n \over N}\,,
\label{theta}
\eeq
that is, the relative chiral rotation of the vacua at $\pm\infty$ is  
proportional to the number of fermions trapped in the ``bag". Our result
(\ref{theta}) is consistent with the well 
known result that the fermion number current in a soliton background can 
be  determined in some cases by topological 
considerations\cite{goldstone}. Note from (\ref{theta}), that in the 
large $N$ limit, $\theta$ (and therefore $\om_1$)
take on non-trivial values only when the number of the trapped fermions 
scales
as a finite fracion of $N$.

As we already mentioned in the introduction, stability of ``bags" 
formed in the NJL model are not stable because of topology. They are 
stabilised by releasing binding energy of the fermions trapped in them.
To see this more explicitly, we calculate now the mass of the ``bag" 
corresponding to (\ref{sigmapi1bs}),(\ref{quantisation1bs}) and 
(\ref{theta}).
The effective action (\ref{effective}) for background fields 
(\ref{sigmapi1bs}) is an ordinary function of the chiral angle 
$\theta$~\footnote{Recall that $\om_1$ is a function of $\theta$ and not 
a free parameter.}. Let us denote 
this action per unit time by $S(\theta)/T$. Then, from (\ref{diagonal}) 
and
(\ref{spectral}) we find   

\beqra
{1\over T}{\pa S\over \pa\theta} = \int{dx\,d\om \over 2\pi}\left\{ 
\left[{i\sigx\over \sqrt{m^2-\om^2}}-(A+D)\right]{\pa \sigx\over 
\pa\theta}\right.
\nonumber\\{}\nonumber\\
\left. +i\left[{\pix\over \sqrt{m^2-\om^2}} - (B+C)\right]{\pa\pix\over 
\pa\theta}\right\}\,.
\label{dsdtheta}
\eeqra
Then as in our analysis of the saddle point condition (which is simply 
the
condition for $ {\pa S\over \pa\theta}=0$) we use (\ref{bofx1final}), 
(\ref{cofx1final}), (\ref{abcd}) and the fact that (\ref{fox}) vanishes
to find

\beqast
{1\over T}{\pa S \over \pa\theta} = i\int{dx\,d\om \over 2\pi} 
{\pax\sigx \pa_{\theta}\pix-\pax\pix \pa_{\theta}\sigx
\over 2\left(\om-\om_1\right)\sqrt{m^2-\om^2} }\,.
\eeqast
Then, (\ref{pisigma}) leads to the factorised expression

\beq
{1\over T}{\pa S\over \pa\theta} = {i\over 4 {\rm sin}^2{\theta\over 
2}}\int\limits_{-\infty}^{\infty}\,dx\,(\si-m)\si'\int_{\cal C'} {d\om 
\over 2\pi}\,{1\over \left(\om-\om_1\right)\sqrt{m^2-\om^2}}
\eeq
where ${\cal C'}$ is the contour in Fig.(2). The space integral is 
immediate
and is essentially fixed by the boundary conditions 
(\ref{boundaryconditions}).
The spectral integral is given by the left hand side of (\ref{arctan}), 
but 
with a generic $\om_1$ given by $\om_1= m{\rm cos} ({\theta\over 2})$. 
Here we
have chosen the particular branch of (\ref{luckily}) that contains all 
the extremal values of $\om_1$. Putting everything together we finally 
arrive at 

\beq
{1\over NT}{\pa S\over \pa\theta} = {m\over 2}\left({n\over N} + 
{\theta\over 
2\pi}\right) {\rm sin}{\theta\over 2}\,.
\label{nigzeret}
\eeq
The zeros of (\ref{nigzeret}) are simply the zeros of (\ref{arctan}), as 
these two equations are one and the same extremum condition. Integrating 
(\ref{nigzeret}) with respect to $\theta$ 
we finally find

\beq
{-1\over NTm} S(\theta)  = \left({n\over N} + {\theta\over 
2\pi}\right) {\rm cos}{\theta\over 2} - {1\over \pi}{\rm sin}{\theta\over 
2}\,.
\label{stheta}
\eeq
Note that (\ref{stheta}) is not manifestly periodic in $\theta$ because 
the Pauli exclusion principle limits $\theta$ to be between $0$ and 
$2\pi$.

The mass of a ``bag" containing $n$ fermions in a single bound state
is given by $-S/T$ evaluated at the appropriate chiral angle 
(\ref{theta}).
We thus find that this mass is simply
\beq
M_n={Nm\over\pi} {\rm sin}{\pi n\over N}
\label{bagmass}
\eeq
in accordance with \cite{shei,andrei}. These ``bags" are stable because 

\beq
{\rm sin}{\pi \left(n_1 + n_2\right)\over N} < {\rm sin}{\pi n_1\over 
N}\,+\,{\rm sin}{\pi n_2\over N}
\label{stablebag}
\eeq
for $n_1, n_2$ less than $N$, such that a ``bag" with $n_1+n_2$ fermions 
cannot decay into two ``bags" each containing a lower number of fermions.

Entrapment of a small number of fermions cannot distort the homogeneous 
vacuum considerably, so we expect that $M_n$ will be roughly the mass of 
$n$ free massive fermions for $n<<N$. As a matter of fact we used this 
expectation to
fix the integration constant in (\ref{bagmass}). For $n<<N$ we have 
$M_n\sim nm [1- {1\over 6}\left({\pi n \over N}\right)^2+\cdots]$, so the 
binding energy released 

\beq
B_n\sim {nm\over 6}\left({\pi n \over N}\right)^2 +\cdots
\label{binding}
\eeq
is indeed very small. However, as the number of fermions trapped in the 
``bag" approaches $N$, $M_n$ vanishes and the fermions release
practically all their rest mass $Nm$ as binding energy, to achieve 
maximum stability\cite{mackenzie}. In a weakly coupled field theory 
containing 
solitons, the mass of these extended objects is a measure of ${1\over 
g^2}$, 
the inverse square of the coupling constant. Here we have ${1\over 
g^2}=N$.
It is amusing to speculate that these maximally stable massless
solitons may teach is something about the strong coupling regime of the 
NJL model.

Note from (\ref{theta}), that the soliton twists all the way around as 
the number of fermions approaches $N$. In this case $\om_1\rightarrow 
-m$, and the pole the resolvent has at $\om=\om_1$ pinches the 
branchpoint $\om=-m$ at the edge of the filled Dirac sea. One may wonder 
whether this enhanced 
singularity is a mathematical artifact, as the bound state simply tries 
to plunge into the filled Dirac sea. But this is clearly not the case. 
Indeed, $\om_1$ is occupied by $N$ fermions 
(in a flavor singlet). Their common spinor wave function must still be
part of the discrete spectrum of the Dirac operator, because the highest 
lying state of the sea at $\om=-m$ is already occupied by a flavor 
singlet 
made of $N$ fermions, sharing a continuum spinor wave function, and 
therefore
Pauli's exclusion principle protects the bound state from ``dissolving" 
into
the sea.

\subsection{Two bound states}

We concluded Subsection 4.2 short of an explicit solution of 
(\ref{sieq2}), namely, short of an explicit expression for the two bound 
state background  fields $\sigx$ and $\pix$. In the following we make the 
eminently reasonable
assumption that such a background exists, and pursue our analysis of its 
saddle point condition as far as we can without having its explicit form 
in hand.

As in the previous subsection, we substitute  (\ref{bofx2final}) and 
(\ref{cofx2final}) into the saddle point 
equations (\ref{saddle1}). We then make use of (\ref{spectral}) to 
eliminate 
the ultraviolate logarithmic divergences and to write
the saddle point conditions as 

\begin{eqnarray}
\int\limits_{\cal C''}\,{d\omega\over 2\pi}~ \left[{i\sigx\over 
\sqrt{m^2-\om^2}} -(A+D)\right] = \nonumber\\{}\nonumber\\
i\int\limits_{\cal C''}\,{d\omega\over 2\pi}~ {K+\om {L\over 
\om^2-\pi^2\left(x\right)} \over 
2\sqrt{m^2-m^2}\left(\om-\om_1\right)\left(\om-\om_2\right)}\nonumber\\{\
rm and}\quad\quad\quad\quad\quad\quad\quad\quad\quad\quad\quad\quad
\nonumber\\
\int\limits_{\cal C''}\,{d\omega\over 2\pi}~ \left[{\pix\over 
\sqrt{m^2-\om^2}} 
-(B+C)\right] = \nonumber\\{}\nonumber\\
\int\limits_{\cal C''}\,{d\omega\over 2\pi}~ {M+\om \si'\left(x\right) 
\over 2\sqrt{m^2-m^2}\left(\om-\om_1\right)\left(\om-\om_2\right)}
\label{saddlepair2bs}
\end{eqnarray}
where ${\cal C''}$ is a contour similar to the contour ${\cal C'}$ in 
Fig.(2) that encircles the additional pole at $\om_2$ as well, 
and $K(x), L(x)$ and $M(x)$ are given by
\beqra 
K(\si, \pi) &=& -\si(\si^2+\pi^2-m^2)+(\om_1+\om_2)\pi'+{1\over 2}\si''
\nonumber\\{}\nonumber\\
L(\si, \pi) &=& (\om_1+\om_2)\left[\si(\si^2+\pi^2-m^2)
-{1\over 2}\si''\right]-\left({\si^2+\pi^2-m^2\over 2} +\om_1\om_2 
+\si^2\right)\pi'\nonumber\\{}\nonumber\\
 &+&{\pi^{'''}\over 4~~}\nonumber\\{\rm and}\nonumber\\
M(\si, \pi) &=& -\pi(\si^2+\pi^2-m^2)-(\om_1+\om_2)\pi'+{1\over 
2}\pi''\,.
\label{KLM}
\eeqra

Note that $K(\si, \pi)$ differs from $-F(\si, \pi)$ in (\ref{fox}) only 
by the additional term $\om_2\pi'$. The expression $L(\si, \pi)$ is the 
residue of the $x$ dependent poles at $\om = \pm\pix$ in the first 
equation in (\ref{saddlepair2bs}). The quantisation conditions 
(\ref{saddlepair2bs}) on $\om_1$ and $\om_2$ cannot be $x$ dependent. 
Therefore 
$L(\si, \pi)$  must vanish as a consistency requirement. As we do not 
have the
explicit expressions of $\sigx$ and $\pix$, we assume from now on that 
$L$ indeed vanishes. This is the only extra assumption we make. Then, 
assuming that the ``bag" traps $n_1$ fermions in $\om_1$ and $n_2$ 
fermions in $\om_2$,  (\ref{saddlepair2bs}) boils down to the simple 
conditions

\beq
I(\om_1)\,=\,I(\om_2)\,=\,0
\label{arctan12}
\eeq
where $I(\om)$ is given in (\ref{arctan}). Therefore,
\beq
\om_1 = m\, {\rm cos}\left(n_1\pi\over N\right)\,,\quad\quad \om_2 = m 
\,{\rm cos}\left(n_2\pi\over N\right)
\label{quantisation12}
\eeq
which are identical in form to single bound state energy levels. From the 
general considerations of \cite{goldstone} we expect that the chiral 
angle
$\theta$ will be proportional to the total number of fermions trapped by 
the ``bag", so (\ref{theta}) must read now

\beq
\theta = - {2\pi \left(n_1 + n_2\right) \over N}\,.
\label{theta2bs}
\eeq
The soliton mass is a function of $m$ and of the chiral angle $\theta$. 
Assuming this function is the same as in the previous case we therefore 
conjecture that the mass of the two bounnd state ``bag" is simply

\beq
M_{n_1,n_2}={Nm\over\pi} {\rm sin}{\pi \left(n_1 + n_2\right) \over N}\,.
\label{bagmass2bs}
\eeq
As far as mass is concerned, such a ``bag" cannot be distinguished from
a single bound state ``bag" containing the same total number $n_1+n_2$ 
of trapped fermions. If our conjecture is true, then such ``bags" are 
stable against decaying into several ``bags" with  
lower numbers of fermions as (\ref{stablebag}) shows.

\pagebreak

\pagebreak

\appendix

\section{Appendix A}
\setcounter{equation}{0}
In this Appendix we provide precise definitions of the Jost functions 
$b_1$ through $c_2$ in terms of their spatial asymptotic behavior and 
derive
the spatial asymptotic behavior of the static Dirac operator Green's 
function.

We concentrate for the moment on the first equation in (\ref{bc}). The 
boundary conditions (\ref{boundaryconditions}) lead to the following 
simple spatial asymptotic behavior 

\beqast
[-\pa_x^2 + m^2-\om^2]\, b(x)=0 
\eeqast
of the homogeneous part of that equation. Thus, solutions of that 
homogeneous equation assume the generic asymptotic form
\beq
b(x,\om)\sim\left\{\begin{array}{cc} M_{+} e^{ikx} + N_{+} e^{-ikx}\,,
\quad\quad & x\rightarrow +\infty \\{}&{}\\ 
 M_{-} e^{ikx} + N_{-} e^{-ikx}\,,\quad\quad 
& x\rightarrow -\infty\end{array}\right.
\label{basymptotic}
\eeq
where
\beq
k(\om)=+\sqrt{\om^2-m^2}\,.
\label{k}
\eeq
On the real $\om$ axis $k(\om)$ is real for $|\om|>m$, which corresponds 
to scattering 
states of (\ref{dirac1}). Bound states of (\ref{dirac1}) reside in the 
domain $|\om|<m$, where $k(\om)=+i\sqrt{m^2-\om^2}=+i\kappa(\om)$ is 
purely imaginary and lies in the upper half plane.
 
The Jost functions $b_1$ and $b_2$ alluded to in Section 2 form a 
particular 
pair of linearly independnt solution of the homogeneous equation 
mentioned above,
specified by their asymptotic behavior. Let the asymptotic amplitudes of 
$b_r(x),\quad (r=1,2)$ in (\ref{basymptotic}) be $M_{r\pm}, N_{r\pm}$. 
The asymptotic form (\ref{basymptotic}) of $b_1(x)$ has by definition 
$M_{1-}=0$, and that of $b_2(x)$ has $N_{2+}=0$. One may summarise our 
definitions of 
$b_1$ and $b_2$, by saying that $b_1$ corresponds to a one dimensional 
scattering situation where the source is at $+\infty$  emitting waves to 
the left (the term $N_{1+}e^{-ikx}$) and that $b_2$ corresponds to a one 
dimensional scattering situation where the source is at $-\infty$  
emitting waves to the right (the term $M_{2-}e^{ikx}$). Note 
also that outside the continuum, $b_1(x)$ decays to the left
while $b_2(x)$ decays to the right.
With these definitions the Wronskian (\ref{wronskian}) becomes
\beq
W_b(+\infty) = -2ik {M_{2+}N_{1+}\over \om+\pi\left(+\infty\right)} = 
-2ik {M_{2-}N_{1-}\over \om+\pi\left(-\infty\right)} = W_b(-\infty)\,.
\label{wronskianb}
\eeq
Therefore, it follows from (\ref{abcd}) and from (\ref{bcrelation}) that 
the entries $A, B, C,$ and $D$ 
in (\ref{diagonal}) have the asymptotic form

\beqra
A(x) &=& -{1\over 2k} \left\{\left[\sigx - ik\, {\rm sgn}x \right]R(k)\, 
e^{2ik|x|}\, + \sigx\right\}\nonumber\\{}\nonumber\\
B(x) &=& {\om+\pix\over -2ik}\left[ 1 + R(k)\, 
e^{2ik|x|}\,\right]\nonumber\\{}\nonumber\\
C(x) &=& {\om-\pix\over 2ik}\left[ 1 + R(k)\, 
e^{2ik|x|}\,\right]\nonumber\\{}\nonumber\\
D(x) &=& -{1\over 2k} \left\{\left[\sigx + ik\, {\rm sgn}x \right]R(k)\, 
e^{2ik|x|}\, + \sigx\right\}
\label{asymptoticabcd}
\eeqra
as $x\rightarrow\pm\infty$, where 
\beq
R(k)={M_{1+}\over N_{1+}} = {N_{2-}\over M_{2-}}
\label{reflection}
\eeq
is the reflection coefficient of the Sturm-Liouville operator in the 
first
equation in (\ref{bc}). The diagonal resolvent of the Dirac operator is 
therefore

\beqra
\langle x\,|-iD^{-1} | x\,\rangle \asymptotic {i\over 2k}
\left(\begin{array}{cc} i\sigx & \omega+\pix 
\\{}&{}\\ -\omega+\pix &  
i\sigx\end{array}\right)&+&\nonumber\\{}\nonumber\\{}\nonumber\\
{iR\left(k\right) e^{2ik,|x|}\over 2 k}
\left(\begin{array}{cc} i\sigx +k\, {\rm sgn}\, x & \omega+\pix 
\\{}&{}\\ -\omega+\pix &  i\sigx -k\, {\rm sgn}\, x 
\end{array}\right)\,&.& 
\label{asymptoticABCD}
\eeqra

\pagebreak

\section{Appendix B}
\setcounter{equation}{0}

Consider the Sturm-Liouville problem
\beq
-\left[p(x)\psi^{'}(x)\right]^{'} + 
\left[V(x)-E\rho(x)\right]\psi(x)\,=\,0\quad\quad,\quad\quad-\infty<x<
\infty\,.
\label{sl}
\eeq

We assume that the ``metric" $p(x)$ does not vanish anywhere and that the  
weight 
function 
$\rho(x)$  is positive everywhere. $E$ is a complex number, called the 
spectral 
parameter.

As in our discussion in the main text and in the previous appendix, let 
$\psi_1(x)$ be the Jost function which decays as $x\rightarrow -\infty$ 
for values of $E$  below the continuum threshold. Similarly, let 
$\psi_2(x)$ be
the Jost function which decays as $x\rightarrow +\infty$. Then, the 
Green's function of the operator in (\ref{sl}) is 

\beq
G(x,y) = 
{\theta\left(x-y\right)\psi_2(x)\psi_1(y)+\theta\left(y-x\right)\psi_2(y)
\psi_1(x)\over W}
\label{slgreens}
\eeq
where 

\beq
W =p(x)\left[\psi_2(x)\psi_1^{'}(x)-\psi_1(x)\psi_2^{'}(x)\right]
\label{slwronskian}
\eeq
is the ($x$ independent) Wronskian of these two functions.  Note that 
(\ref{slgreens}) decays (at a rate dictated by the Jost functions) as 
either one of its argument diverges in absolute value, holding the other 
one finite, as
long as $E$ does not hit one of the eigenvalues of the Sturm-Liouville 
operator.

As in the main text the diagonal resolvent $R(x) = G(x,x)$ is defined as 

\beq 
R(x) = {1\over 2}\lim_{\epsilon\rightarrow 0}\left[G(x,x+\epsilon) + 
G(x+\epsilon, x)\right] = {\psi_1(x)\psi_2(x)\over W}\,.
\label{diagonalres}
\eeq

We then use (\ref{slwronskian}) and (\ref{diagonalres}) to show that
\beq
{\psi^{'}_1 \over \psi_1} = {pR^{'} + 1 \over 2pR}\quad\quad,\quad\quad 
{\psi^{'}_2 \over \psi_2} = {pR^{'} - 1 \over 2pR}\,.
\label{psi12}
\eeq

Finally, using (\ref{sl}) and (\ref{psi12}) we find 

\beq
(pR^{'})^{'} = 2(V-E\rho)R + {\left(pR^{'}\right)^2 - 1 \over 2pR}
\label{pregd}
\eeq
and

\beq
[p(pR^{'})^{'}]^{'} = [2p(V-E\rho)R]^{'} +2p(V-E\rho)R^{'}\,. 
\label{lineargd}
\eeq

Note that the non-linearity of (\ref{pregd}) in $R$ has disappeared after 
one more differentiation with respect to $x$.

Multiplying (\ref{pregd}) through by $2R$ we find
\beq
-2pR(pR^{'})^{'} + \left(pR^{'}\right)^2 + 4pR^{2}(V-\rho E) = 1
\label{slgd}
\eeq
which is the Gel'fand-Dikii equation \cite{gd}. Eq. (\ref{lineargd}) is 
the 
linear form of the Gel'fand-Dikii equation we use in the text (Eq. 
(\ref{gdblinear} .) 

The quantities corresponding to the discussion of the Dirac operator in 
the text are 
\beqra
p(x) &=& \rho(x) =  {1\over \omega + \pix}\quad\quad,\quad\quad 
E=\omega^2
\nonumber\\{\rm and}\nonumber\\
V(x) &=&  {1\over \omega + \pix}\cdot\left[\sigx^2+\pix^2-\si'(x) + 
{\sigx\pi'(x)\over \omega+\pix}\right]\,.
\label{specific}
\eeqra

The Gel'fand-Dikii equation (\ref{slgd}) then reads
\beqra
-2{B(x)\over \omega+\pix}\pax\left[{\pax B(x)\over \omega+\pix}\right] +
\left[{\pax B(x)\over \omega+\pix}\right]^2 
+~~~~~~~~~~~~~~~~~~\nonumber\\{}\nonumber\\  \left[{2B(x) \over 
\omega+\pix}
\right]^2\left[\sigx^2+\pix^2-\si'(x)-\omega^2+{\sigx\pi'(x)\over 
\omega+\pix}\right] = 1\,.
\label{gdb}
\eeqra

Strictly speaking, Sturm-Liouville theory requires that 
$p(x)=\rho(x)={1\over \omega + \pix} > 0\,.$ Our solution for $\pix$ 
turns 
out to be bounded, so 
all formulae are valid a posteriori for large positive $\omega$. Such a 
restriction on $\omega$, though mathematically required, is unphysical. 

Note however, that because of the relation (\ref{bcrelation}), we may 
view 
$C(x)$ 
as a continuation of $B(x)$ to large negative $\omega$.

An important application of the Gel'fand-Dikii identities 
(\ref{lineargd}),  (\ref{slgd}) is that they generate an asymptotic 
expansion\cite{gd} of $R$ in
negative odd powers of $\sqrt{E}$. The explicit $\omega$  (and therefore 
$E$) dependence of  our specific $p(x), \rho(x)$ and $V(x)$ complicates 
this expansion.

{\bf Acknowledgements}~~~ We would like to thank N. Andrei for valuable 
discussions, and J. Polchinski for reminding us of \cite{goldstone}.
This work was partly supported by the National Science Foundation under 
Grant
No. PHY89-04035.

\end{document}

%% file: epsfig.tex
\newread\epsffilein    
\newif\ifepsffileok    
\newif\ifepsfbbfound   
\newif\ifepsfverbose   
\newdimen\epsfxsize    
\newdimen\epsfysize    
\newdimen\epsftsize    
\newdimen\epsfrsize    
\newdimen\epsftmp      
\newdimen\pspoints     
\def\epsfbox#1{\global\def\epsfllx{72}\global\def\epsflly{72}%
   \global\def\epsfurx{540}\global\def\epsfury{720}%
   \def\lbracket{[}\def\testit{#1}\ifx\testit\lbracket
   \let\next=\epsfgetlitbb\else\let\next=\epsfnormal\fi\next{#1}}%
\def\epsfgetlitbb#1#2 #3 #4 #5]#6{\epsfgrab #2 #3 #4 #5 .\\%
   \epsfsetgraph{#6}}%
\def\epsfnormal#1{\epsfgetbb{#1}\epsfsetgraph{#1}}%
\def\epsfgetbb#1{%
\openin\epsffilein=#1
\ifeof\epsffilein\errmessage{I couldn't open #1, will ignore it}\else
   {\epsffileoktrue \chardef\other=12
    \def\do##1{\catcode`##1=\other}\dospecials \catcode`\ =10
    \loop
       \read\epsffilein to \epsffileline
       \ifeof\epsffilein\epsffileokfalse\else
          \expandafter\epsfaux\epsffileline:. \\%
       \fi
   \ifepsffileok\repeat
   \ifepsfbbfound\else
    \ifepsfverbose\message{No bounding box comment in #1; using defaults}\fi\fi
   }\closein\epsffilein\fi}%
\def\epsfclipstring{}
\def\epsfsetgraph#1{%
   \epsfrsize=\epsfury\pspoints
   \advance\epsfrsize by-\epsflly\pspoints
   \epsftsize=\epsfurx\pspoints
   \advance\epsftsize by-\epsfllx\pspoints
   \epsfxsize\epsfsize\epsftsize\epsfrsize
   \ifnum\epsfxsize=0 \ifnum\epsfysize=0
      \epsfxsize=\epsftsize \epsfysize=\epsfrsize
      \epsfrsize=0pt
     \else\epsftmp=\epsftsize \divide\epsftmp\epsfrsize
       \epsfxsize=\epsfysize \multiply\epsfxsize\epsftmp
       \multiply\epsftmp\epsfrsize \advance\epsftsize-\epsftmp
       \epsftmp=\epsfysize
       \loop \advance\epsftsize\epsftsize \divide\epsftmp 2
       \ifnum\epsftmp>0
          \ifnum\epsftsize<\epsfrsize\else
             \advance\epsftsize-\epsfrsize \advance\epsfxsize\epsftmp \fi
       \repeat
       \epsfrsize=0pt
     \fi
   \else \ifnum\epsfysize=0
     \epsftmp=\epsfrsize \divide\epsftmp\epsftsize
     \epsfysize=\epsfxsize \multiply\epsfysize\epsftmp   
     \multiply\epsftmp\epsftsize \advance\epsfrsize-\epsftmp
     \epsftmp=\epsfxsize
     \loop \advance\epsfrsize\epsfrsize \divide\epsftmp 2
     \ifnum\epsftmp>0
        \ifnum\epsfrsize<\epsftsize\else
           \advance\epsfrsize-\epsftsize \advance\epsfysize\epsftmp \fi
     \repeat
     \epsfrsize=0pt
    \else
     \epsfrsize=\epsfysize
    \fi
   \fi
   \ifepsfverbose\message{#1: width=\the\epsfxsize, height=\the\epsfysize}\fi
   \epsftmp=10\epsfxsize \divide\epsftmp\pspoints
   \vbox to\epsfysize{\vfil\hbox to\epsfxsize{%
      \ifnum\epsfrsize=0\relax
        \includegraphics{#1}%
      \else
        \epsfrsize=10\epsfysize \divide\epsfrsize\pspoints
        \includegraphics{#1}%
      \fi
      \hfil}}%
\global\epsfxsize=0pt\global\epsfysize=0pt}%
{\catcode`\%=12 
\global\let\epsfpercent=
\long\def\epsfaux#1#2:#3\\{\ifx#1\epsfpercent
   \def\testit{#2}\ifx\testit\epsfbblit
      \epsfgrab #3 . . . \\%
      \epsffileokfalse
      \global\epsfbbfoundtrue
   \fi\else\ifx#1\par\else\epsffileokfalse\fi\fi}%
\def\epsfempty{}%
\def\epsfgrab #1 #2 #3 #4 #5\\{%
\global\def\epsfllx{#1}\ifx\epsfllx\epsfempty
      \epsfgrab #2 #3 #4 #5 .\\\else
   \global\def\epsflly{#2}%
   \global\def\epsfurx{#3}\global\def\epsfury{#4}\fi}%
\def\epsfsize#1#2{\epsfxsize}